\documentclass[aps,prb,preprint,superscriptaddress,longbibliography]{revtex4-1}
\usepackage{amsmath}
\usepackage{amssymb}
\usepackage{color}
\usepackage{epsfig}
\usepackage{graphicx}
\usepackage{longtable}
\usepackage{natbib}
\usepackage{revsymb}
\usepackage{threeparttable}
\usepackage{txfonts}
\usepackage{xr}
\usepackage{verbatim}
\usepackage{threeparttable}
\usepackage{bm}
\usepackage{braket}
\usepackage{multirow}
\usepackage{tabularx}
\graphicspath{{figs/}}

\begin{document}

\title{Synergy of Rashba and Topological Effects for High-Performance Bismuth-Based Thermoelectrics}

\author{Lei Peng}
\affiliation{School of Information Science and Technology, Department of Optical Science and Engineering, Fudan University, Shanghai 200433, China}
\affiliation{Key Laboratory for Computational Physical Sciences (MOE), Institute of Computational Physical Sciences and Department of Physics, Fudan University, Shanghai 200433, China}

\author{Ruixiao Lian}
\affiliation{School of Information Science and Technology, Department of Optical Science and Engineering, Fudan University, Shanghai 200433, China}

\author{Hongyu Chen}
\affiliation{School of Information Science and Technology, Department of Optical Science and Engineering, Fudan University, Shanghai 200433, China}

\author{Ben Li}
\affiliation{School of Information Science and Technology, Department of Optical Science and Engineering, Fudan University, Shanghai 200433, China}

\author{Yu Wu}
\affiliation{Advanced Thermal Management Technology and Functional Materials Laboratory, Ministry of Education Key Laboratory of NSLSCS, School of Energy and Mechanical Engineering, Nanjing Normal University, Nanjing 210023, P. R. China}

\author{Yuxiang Zheng} \email{yxzheng@fudan.edu.cn}
\affiliation{School of Information Science and Technology, Department of Optical Science and Engineering, Fudan University, Shanghai 200433, China}
\affiliation{Yiwu Research Institute of Fudan University, Yiwu City, Zhejiang 322000, China}

\author{Hao Zhang} \email{zhangh@fudan.edu.cn}
\affiliation{School of Information Science and Technology, Department of Optical Science and Engineering, Fudan University, Shanghai 200433, China}
\affiliation{Yiwu Research Institute of Fudan University, Yiwu City, Zhejiang 322000, China}

\begin{abstract}
	Band convergence is a key strategy for enhancing thermoelectric (TE) performance. Herein, we demonstrate a promising approach to enhance band convergence through inducing Rashba splitting in topological insulators. Theoretically designed Janus $\beta$-Bi$_2$Se$_2$Te and $\beta$-Bi$_2$Te$_2$Se exhibited inherent topological band inversion and Rashba splitting due to the strong spin-orbit coupling (SOC) with broken inversion symmetry. These characteristics synergistically improve band convergence, leading to a substantially enhanced power factor. Meanwhile, the Janus structural asymmetry suppresses lattice thermal conductivity. Consequently, Janus structures achieve boosted TE performance, especially for $\beta$-Bi$_2$Se$_2$Te, peaking figure of merit ($zT$) of 2.82. This work establishes a new framework for designing Janus topological compounds with high TE performance by the synergistic effect of Rashba splitting and band inversion.
\end{abstract}

\maketitle

\section{Introduction}

Thermoelectric (TE) materials offer a sustainable solution to the global energy crisis by enabling the direct and reversible conversion between waste heat and electrical energy.\cite{xiao_seeking_2020,snyder_complex_2008,zhang_thermoelectric_2015}
The efficiency of TE materials is quantified by the dimensionless figure of merit, $zT = S^2\sigma T/\kappa$, where $S$ is the Seebeck coefficient, $\sigma$ is the electrical conductivity, $T$ is the absolute temperature, and $\kappa$ is the total thermal conductivity, comprising electronic ($\kappa_e$) and lattice ($\kappa_{L}$) contributions. 
The primary challenge in enhancing $zT$ lies in navigating the intricate and often conflicting interdependencies among these parameters. 
In particular, achieving a high power factor, defined as $PF = S^2\sigma$, remains challenging due to the inverse relationship between $S$ and $\sigma$.
Consequently, discovering novel materials and mechanisms that can simultaneously optimize electronic transport while suppressing thermal transport remains a central goal in materials science.\cite{he_advances_2017,shi_advanced_2020}

A cornerstone strategy for augmenting the power factor is band convergence, which involves aligning multiple electronic band valleys at or near the Fermi level.
This approach effectively increases the band degeneracy, leading to a larger density of states effective mass. Such an enhancement boosts the Seebeck coefficient without significantly compromising carrier mobility ($\mu$), thereby decoupling the typically inverse relationship between $S$ and $\sigma$.\cite{pei_convergence_2011,shi_global_2024}
This principle has been instrumental in achieving benchmark $zT$ values in classic thermoelectric materials like PbTe-based alloys, where the convergence of up to 12 valleys has been realized.\cite{pei_convergence_2011}
However, rationally designing material systems with inherent multi-valley band structures or engineering such features through targeted structural modifications presents an ongoing research frontier.

Bismuth-based chalcogenides, such as Bi$_2$Te$_3$ and its alloys, are canonical high-performance thermoelectric materials, renowned for their favorable electronic properties and low lattice thermal conductivity.\cite{witting_thermoelectric_2019,pei_bi2te3-based_2020,hong_fundamental_2018}
Many of these compounds are also topological insulators (TIs), characterized by strong spin-orbit coupling (SOC) that induces an inversion of bulk bands near the Fermi level.\cite{zhang_topological_2009,cai_independence_2018,ren_large_2010}
This intrinsic electronic feature offers a unique platform for thermoelectric enhancement.\cite{heremans_tetradymites_2017}
It is hypothesized that by judiciously breaking the spatial inversion symmetry in these TI-based materials, new avenues for band engineering can be unlocked. The breaking of this symmetry is known to induce the Rashba effect, which lifts spin degeneracy and splits the energy bands.
This work proposes a novel design concept: to leverage the synergy of SOC induced rashba and topological effects. This combination is posited to enrich the band structure, creating a convergence of multiple valleys near the Fermi level and thereby enhancing thermoelectric performance. To realize this, two non-centrosymmetric Janus compounds, which we term $\beta$-phases ($\beta$-Bi$_2$Se$_2$Te and $\beta$-Bi$_2$Te$_2$Se), were computationally designed based on their known centrosymmetric prototypes, hereafter denoted as $\alpha$-phases ($\alpha$-Bi$_2$TeSe$_2$ and $\alpha$-Bi$_2$SeTe$_2$).\cite{nakajima_crystal_1963}

In this work, the stability, electronic topology, and thermoelectric performance of the designed $\beta$-phase Janus structures ($\beta$-Bi$_2$Se$_2$Te, $\beta$-Bi$_2$Te$_2$Se) and their $\alpha$-phase prototypes are systematically investigated via first-principles calculations. Phonon spectra and mechanical property analyses confirm dynamical stability, while Wilson loop $Z_2$ invariants and surface states verify their TI nature. Crucially, the $\beta$-phase Janus systems exhibit enhanced band degeneracy from Rashba splitting, superior power factor, and ultralow $\kappa_{L}$, synergistically boosting $zT$ beyond 1. This study not only identifies promising new candidates for thermoelectric applications but also demonstrates a powerful design strategy that leverages the synergy of SOC-driven Rashba and topological effects to achieve high-efficiency thermoelectrics.

\section{Methods}

Density functional theory (DFT) calculations within the plane-wave expansion were performed using the Quantum ESPRESSO package.\cite{giannozzi_quantum_2009} 
The generalized gradient approximation (GGA) of Perdew-Burke-Ernzerhof (PBE) with relativistic norm-conserving pseudopotentials were used to describe the exchange-correlation functional.\cite{perdew_generalized_1996,van_setten_pseudodojo_2018} 
The kinetic energy cutoff was set to 80 Ry and and a uniform $11\times11\times11$ $k$-point mesh was used to sample the Brillouin zone. 
The maximally localized Wannier functions were constructed using the Wannier90 code.\cite{mostofi_updated_2014} 
The edge state were calculated by the Green's function method on a tight binding model based on MLWFs within the wanniertools package.\cite{wu_wanniertools_2018}

The electrical transport properties were evaluated using the AMSET code in conjunction with the VASP software package to compute scattering rates and carrier mobilities.\cite{ganose_efficient_2021} The momentum space grid were interpolated to resolutions of $41\times41\times41$ for $\alpha$-phases and $43\times43\times43$ for $\beta$-phases. 
The analysis accounted for key scattering mechanisms, including acoustic deformation potential (ADP) scattering, polar optical phonon (POP) scattering, and ionized impurity (IMP) scattering. Carrier relaxation times were subsequently determined and integrated with the BoltzTraP2 software package to solve the Boltzmann transport equation, enabling the derivation of detailed electrical transport properties.\cite{madsen_boltztrap2_2018}

The ground-state phonon spectrum was computed using the finite displacement method implemented in the Phonopy software. To accurately capture interatomic interactions, calculations were performed on a $4\times4\times1$ supercell derived from the conventional unit cell of the Bi$_2$X$_2$Y compound through non-diagonal cell expansion.\cite{togo_first-principles_2023-1,togo_implementation_2023} Finite-temperature phonon spectra and third-order force constants were obtained using the TDEP package, employing a self-consistent sampling approach.\cite{knoop_tdep_2024,shulumba_intrinsic_2017} The cutoff radii for the extraction of second- and third-order force constants were set to 8 \AA{} and 7 \AA{}, respectively. Thermal transport properties were subsequently evaluated using an iterative solution to the phonon Boltzmann transport equation, with the q-point grid resolution set to $15\times15\times15$. 

The low-energy effective model was constructed using the \( \mathbf{k} \cdot \mathbf{p} \) perturbation method [28], enabling the electronic band structure to be derived near high-symmetry points in \(\mathbf{k}\)-space through invariant theory. Considering a perturbation of the form \(\mathbf{k} = \mathbf{k}_0 + \delta \mathbf{k}\) in the single-electron Schrödinger equation, the system is described by:

\begin{equation}
	\begin{aligned}
	&\left\{ \left[ \frac{\mathbf{p}^2}{2m} + \frac{\hbar}{m} \mathbf{k}_0 \cdot \mathbf{p} + \frac{\hbar^2 \mathbf{k}_0^2}{2m} + V(\mathbf{r}) \right] + \frac{\hbar}{m} \delta \mathbf{k} \cdot \mathbf{p} \right\} u_n(\mathbf{k}_0, \mathbf{r}) \\
	&= \left[ E_n(\mathbf{k}_0) + \frac{\hbar^2 (\mathbf{k}_0^2 - \mathbf{k}^2)}{2m} \right] u_n(\mathbf{k}_0, \mathbf{r}).
	\end{aligned}
\end{equation}
The term \((\hbar /m) \delta \mathbf{k} \cdot \mathbf{p}\) is treated as a perturbation to the Hamiltonian $H_{\mathbf{k} \cdot \mathbf{p}}$ at the point \(\mathbf{k}_0\). By considering only the first-order and second-order perturbative corrections, the eigenvalue is expressed as:

\begin{equation}
    \begin{aligned}
    E_n(\mathbf{k}) = & E_n(\mathbf{k}_0) + \frac{\hbar^2 (\mathbf{k}_0^2 - \mathbf{k}^2)}{2m} \\
    &+ \frac{\hbar}{m} \delta \mathbf{k} \cdot \langle u_n | \mathbf{p} | u_n \rangle + \frac{\hbar^2}{m^2} \sum_l \frac{\left| \langle u_n | \mathbf{p} | u_l \rangle \cdot \delta \mathbf{k} \right|^2}{E_n(\mathbf{k}_0) - E_l(\mathbf{k}_0)}.
    \end{aligned}
\end{equation}
The SOC effect can also be incorporated, yielding the $H_{\mathbf{k} \cdot \mathbf{p}}$ perturbation Hamiltonian: 
\begin{equation}
	H' = H_{\mathbf{k} \cdot \mathbf{p}} + H_{SO}.
\end{equation}
Based on invariant theory, the matrix form of the Hamiltonian comprises $k$-invariant terms within the matrix elements, constrained by the symmetry operations of the system. The $H_{\mathbf{k} \cdot \mathbf{p}}$terms are intrinsically linked to the symmetry operator $\hat{R}$ that preserves the structural symmetry. Under the $\hat{R}$, the Hamiltonian satisfies the relation:
\begin{equation}
D(\hat{R}) H(\mathbf{k}) D(\hat{R})^{-1} = H(\hat{R}\mathbf{k}),
\end{equation}
where \(D(\hat{R})\) is the representation matrix corresponding to the $\hat{R}$ acting on the basis vectors. This representation satisfies:
\begin{equation}
\hat{R} \psi = D(\hat{R}) \psi.
\end{equation}

Using this framework, the low-energy effective Hamiltonian of the $H_{\mathbf{k} \cdot \mathbf{p}}$ model can be systematically derived. The construction of the Hamiltonian model can be streamlined by tools such as DFT2kp, with the model parameters extracted by fitting to values obtained from first-principles calculations.\cite{varjas_qsymm_2018,v_cassiano_dft2kp_2024,zhang_vasp2kp_2023-1}

\section{Results and Discussion}

\subsection{Structures and stability}

\begin{figure}[ht!]
	\centering
	\includegraphics[width=0.9\linewidth]{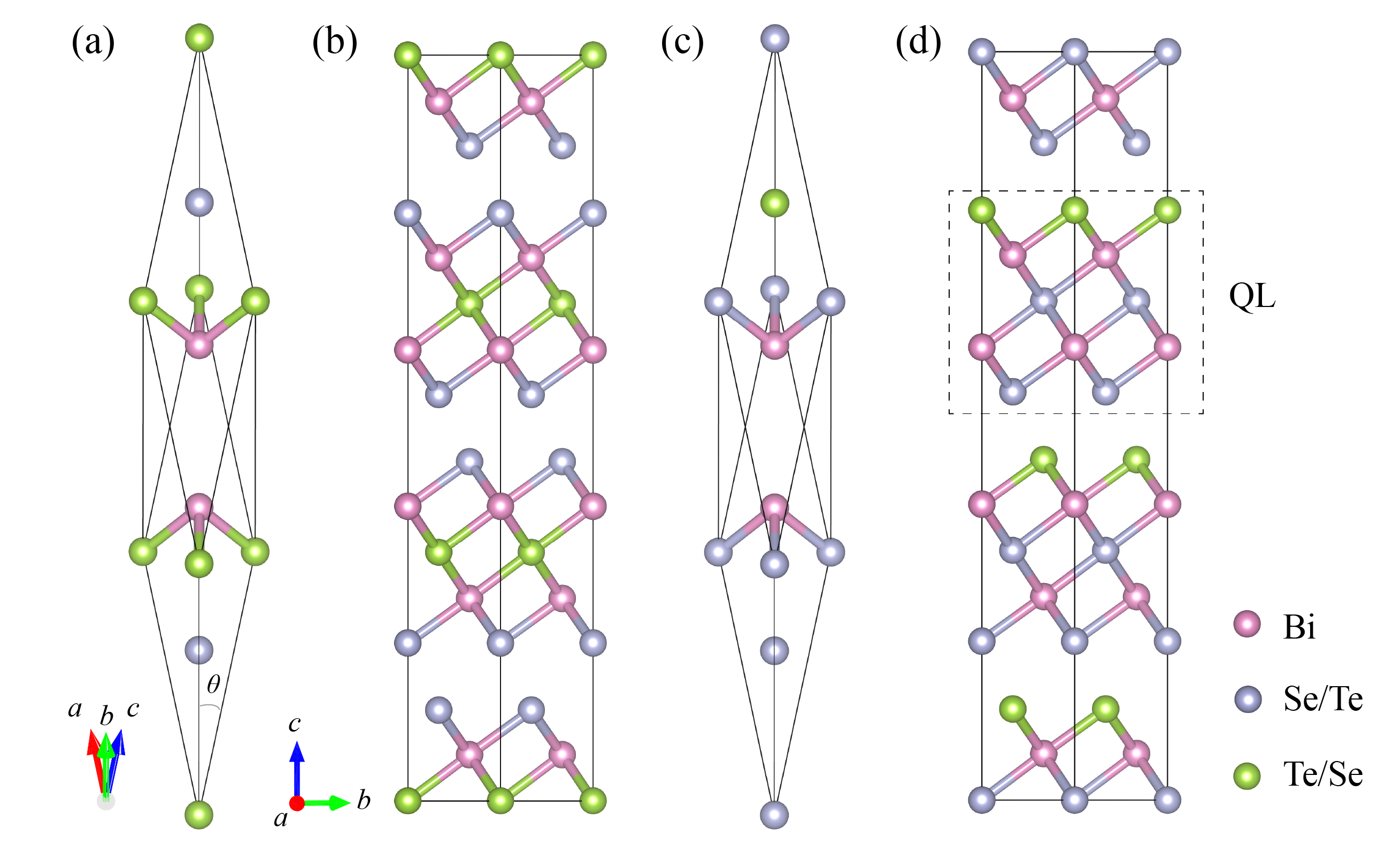}
	\caption{The unit cells and conventional cells of $\alpha$-Bi$_2$TeSe$_2$/$\alpha$-Bi$_2$SeTe$_2$ (a, b) and $\beta$-Bi$_2$Se$_2$Te/$\beta$-Bi$_2$Te$_2$Se (c, d). Dashed boxes indicate the quintuple-layer.}
	\label{fig:01-structure}
\end{figure}

The crystal structures of bulk Bi$_2$X$_2$Y (X, Y = Te, Se) are illustrated in Fig.~\ref{fig:01-structure}, resembling the quintuple-layer (QL) structure of Bi$_2$Te$_3$.
Bi$_2$X$_2$Y exhibits two distinct symmetry types, depending on whether Y or X atoms occupy the central inversion positions. 
Specifically, $\alpha$-phases, $\alpha$-Bi$_2$TeSe$_2$ and $\alpha$-Bi$_2$SeTe$_2$ (Fig.~\ref{fig:01-structure}a, b) possess centrosymmetric $R\bar{3}m$ symmetry, similar to Bi$_2$Te$_3$. In contrast, their counterparts, the $\beta$-phases $\beta$-Bi$_2$Se$_2$Te and $\beta$-Bi$_2$Te$_2$Se (Fig.~\ref{fig:01-structure}c, d), exhibit non-centrosymmetric $R3m$ symmetry, which is characteristic of their Janus structures.
Previous computational studies have demonstrated that the lattice parameters of tetradymite-like materials are sensitive to the choice of exchange-correlation functional and van der Waals interactions, while the electronic and topological properties are less affected under identical lattice parameters.\cite{cao_rhombohedral_2018,cheng_effects_2014}
Accordingly, this study employs experimental lattice parameters for calculations.\cite{nakajima_crystal_1963} Although experimental lattice parameters for the $\beta$-phases are rarely reported, their optimized lattice parameters at the same computational level closely resemble those of the $\alpha$-phase prototypes. 
Therefore, the lattice parameters of 10.05~\AA{} and 10.26~\AA{} are adopted for $\alpha$-Bi$_2$TeSe$_2$/$\beta$-Bi$_2$Se$_2$Te and $\alpha$-Bi$_2$SeTe$_2$/$\beta$-Bi$_2$Te$_2$Se, respectively. The relevant structural parameters are summarized in Table~\ref{tab:table1}. 

\begin{figure}[ht!]
	\centering
	\includegraphics[width=0.9\linewidth]{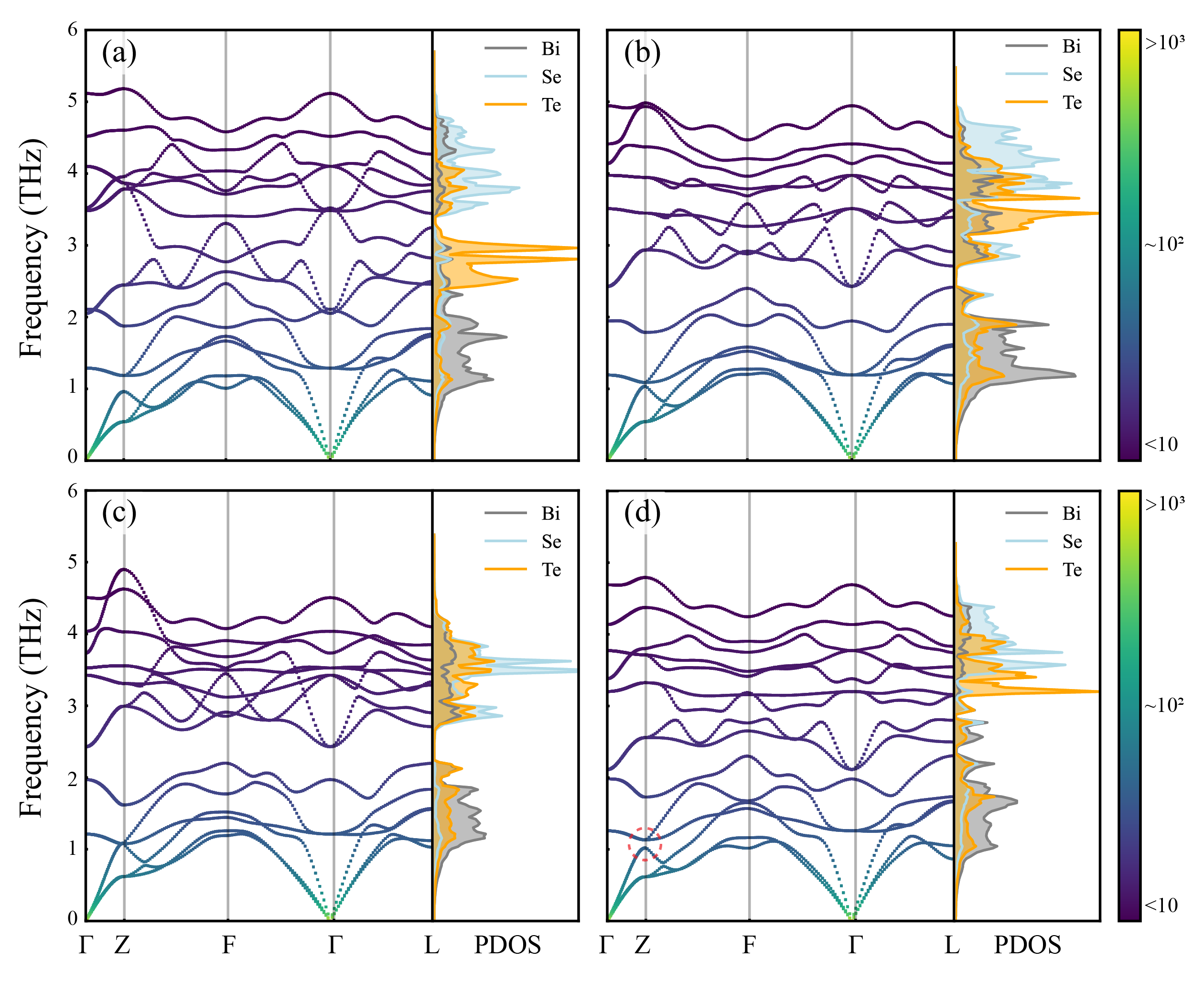}
	\caption{The phonon dispersion with occupancy at 300 K and partial density of states for (a) $\alpha$-Bi$_2$TeSe$_2$, (b) $\beta$-Bi$_2$Se$_2$Te, (c) $\alpha$-Bi$_2$SeTe$_2$, and (d) $\beta$-Bi$_2$Te$_2$Se. The red circle in part d means avoided crossing between the acoustic and optical modes.}
	\label{fig:02-phonondos}
\end{figure}

The ground-state phonon dispersion and partial density of states for Bi$_2$X$_2$Y are depicted in Fig.~\ref{fig:02-phonondos}. The weight distribution is derived from the phonon Bose-Einstein distribution function evaluated at 300 K. The absence of imaginary frequencies in all four structures demonstrates their dynamic stability at low temperatures. Furthermore, the acoustic phonons exhibit a linear dispersion relation in the vicinity of the $\Gamma$ point.
At room temperature, the phonon occupation numbers exceed unity across nearly all frequencies, with particularly high phonon occupancies observed in the low-frequency regime. It is evident that structures with the same stoichiometric ratio, namely $\alpha$-Bi$_2$TeSe$_2$ and $\beta$-Bi$_2$Se$_2$Te (Fig.~\ref{fig:02-phonondos}a, b), as well as $\alpha$-Bi$_2$SeTe$_2$ and $\beta$-Bi$_2$Te$_2$Se (Fig.~\ref{fig:02-phonondos}c, d), exhibit similar maximum frequencies, which decrease as the atomic weight of the constituent elements increases.
The phonon partial density of states reveal that the low-frequency acoustic phonons are predominantly contributed by the heavier Bi and Te atoms, while the vibrational modes of Se atoms are concentrated around 4 THz. Furthermore, the phonon dispersion and density of states exhibit a strong dependence on the atomic configuration. In $\alpha$-Bi$_2$TeSe$_2$, the Te atoms display a distinct density of states distribution compared to the other three materials, with a pronounced contribution in the intermediate frequency range (2–3 THz). This behavior arises from the heavier mass of Te atoms and their central inversion symmetry position, which subjects them to stronger covalent bonding interactions. The localized phonons in this region are expected to participate more actively in acoustic phonon scattering, thereby contributing to a reduction in thermal conductivity.
In contrast, $\alpha$-Bi$_2$SeTe$_2$ exhibits a pronounced phonon band gap in the intermediate frequency range, with the vibrational modes of Te and Se atoms predominantly concentrated in the high-frequency region. This behavior can be attributed to the Te atoms being located on the outer layers of the quintuple atomic layers, where they experience relatively weaker interlayer van der Waals interactions, leading to higher vibrational frequencies. A similar analysis applies to $\beta$-phase Bi$_2$Se$_2$Te and Bi$_2$Te$_2$Se. Overall, the relatively low phonon frequencies in the Bi$_2$X$_2$Y systems are expected to result in lower lattice thermal conductivity.

To further assess the energetic stability, the cohesive energy was calculated at the same computational level using the expression $E_{\text{coh}} = (E_{\text{total}} - nE_{\text{Bi}} - mE_{\text{Te}} - lE_{\text{Se}})/5$, where $E_{\text{total}}$ represents the total energy of the unit cell, and $E_{\text{Bi}}$, $E_{\text{Te}}$, and $E_{\text{Se}}$ denote the energies of individual Bi, Te, and Se atoms in their crystalline states, with $n$, $m$, and $l$ being the respective numbers of Bi, Te, and Se atoms in Bi$_2$X$_2$Y. The calculated cohesive energies are $-2.31$ eV/atom for $\alpha$-Bi$_2$TeSe$_2$, $-2.35$ eV/atom for $\beta$-Bi$_2$Se$_2$Te, $-1.98$ eV/atom for $\alpha$-Bi$_2$SeTe$_2$, and $-1.95$ eV/atom for $\beta$-Bi$_2$Te$_2$Se. All values are negative, and the Janus structures exhibit cohesive energies comparable to those of their prototype counterparts, indicating that all four configurations are energetically stable.

The mechanical properties of Bi$_2$X$_2$Y were systematically investigated, with the relevant parameters summarized in Table S1. The Bi$_2$X$_2$Y crystallizes in a trigonal structure characterized by six independent second-order elastic constants $C_{ij}$. The mechanical stability of these structures was evaluated using the Born-Huang criteria, expressed as:\cite{mouhat_necessary_2014,peng_stability_2017}
\begin{equation}
	\begin{aligned}
		&C_{11} - C_{12} > 0, \\
		&(C_{11} - C_{12})C_{44} - 2C_{14}^2 > 0, \\
		&(C_{11} + C_{12})C_{33} - 2C_{13}^2 > 0.
	\end{aligned}
\end{equation}
All four Bi$_2$X$_2$Y structures satisfy these conditions, confirming their mechanical stability. Additionally, the average bulk modulus, shear modulus, and Young’s modulus were calculated for the four materials (Table S1). The results reveal that their mechanical properties are highly similar, primarily due to the consistent structural features of the Bi$_2$X$_2$Y systems. The three-dimensional spatial distribution of Young’s modulus is illustrated in Fig. S1, showing similar patterns across the four materials. Similarly, other mechanical properties, such as the shear modulus, also exhibit analogous distributions.

Experimentally, the distinct configurations of Bi$_2$X$_2$Y can be identified using infrared (IR) and Raman spectroscopy. Here, we provide a brief analysis of the differences in vibrational modes among the Bi$_2$X$_2$Y structures. For $\alpha$-Bi$_2$TeSe$_2$ and $\alpha$-Bi$_2$SeTe$_2$, which exhibit inversion symmetry, the irreducible representations of their phonon vibrational modes at the $\Gamma$ point in momentum space can be classified by parity (odd or even). In contrast, such parity-based classification is not applicable for $\beta$-Bi$_2$Se$_2$Te and $\beta$-Bi$_2$Te$_2$Se. The point group of $\alpha$-Bi$_2$TeSe$_2$ and $\alpha$-Bi$_2$SeTe$_2$ at the $\Gamma$ point is D$_{3d}$, and the irreducible representations of their optical phonons are given by,\cite{chis_vibrations_2012}
\begin{equation}
	\begin{aligned}
	\Gamma_{\text{vib}} = 2E_{\text{g}}(\text{R}) \oplus 2E_{\text{u}}(\text{I}) \oplus 2A_{2\text{u}}(\text{I}) \oplus 2A_{1\text{g}}(\text{R})
	\end{aligned}
\end{equation}
Based on selection rules, the odd-parity modes ($E_{\text{u}}$ and $A_{2\text{u}}$) are IR-active, while the even-parity modes ($E_{\text{g}}$ and $A_{1\text{g}}$) are Raman-active. For $\beta$-Bi$_2$Se$_2$Te and $\beta$-Bi$_2$Te$_2$Se, the point group at the $\Gamma$ point is C$_{3v}$, which comprises three irreducible representations. The irreducible representation of their optical phonons is given by,

\begin{equation}
	\begin{aligned}
		\Gamma_{\text{vib}} = 4E(\text{R} + \text{I}) \oplus 4A_1(\text{R} + \text{I})
	\end{aligned}
\end{equation}
where $E$ and $A_1$ correspond to the two-dimensional and one-dimensional irreducible representations associated with the LO (TO) and ZO phonon modes, respectively. Since the C$_{3v}$ symmetry group lacks inversion symmetry, the mutual exclusion principle does not apply to any of the active modes. Consequently, all basis functions belonging to the $E$ and $A_1$ irreducible representations include both linear (IR-active) and bilinear (Raman-active) terms. As a result, the phonon modes of $\beta$-phases exhibit both IR and Raman activity.

\begin{table}[ht!]
    \caption{The lattice parameters ($a$, \AA{}), angle ($\theta$, °), DFT band gap ($E^{DFT}_{g}$, eV) and DFT+SOC band gap ($E^{DFT+SOC}_{g}$, eV) of Bi$_2$X$_2$Y. Reported values are given in parentheses.}
	\setlength{\tabcolsep}{3.8mm}{
    \begin{tabular}{cccccccc}
		\hline
		Materials                       & Space group                  & $a$     & $\theta$ & $E^{DFT}_{g}$ & $E^{DFT+SOC}_{g}$  \\ \hline
		\multirow{2}{*}{$\alpha$-Bi$_2$TeSe$_2$} & \multirow{2}{*}{$R\bar{3}m$} & 10.05   & 24.24    & 0.51           & 0.14           \\
										&                              & (9.97, 10.04)\cite{ma_geometric_2018-1,wang_ternary_2011}  & (24.37, 24.20)  & —             & (0.12, 0.17)       \\
		$\beta$-Bi$_2$Se$_2$Te                  & $R3m$                        & 10.05   & 10.89    & 0.21          & 0.27                 \\
		\multirow{2}{*}{$\alpha$-Bi$_2$SeTe$_2$} & \multirow{2}{*}{$R\bar{3}m$} & 10.26   & 24.09    & 0.06          & 0.25                 \\
										&                              & (10.38, 10.26)\cite{ma_geometric_2018-1,wang_ternary_2011} & (23.70, 24.09)  & —             & (0.27, 0.28)            \\
		$\beta$-Bi$_2$Te$_2$Se                  & $R3m$                        & 10.26   & 24.09    & 0.38          & 0.17                  \\ \hline
		\end{tabular}}
    \label{tab:table1}
\end{table}

\subsection{Electronic and topological properties}

\begin{figure}[ht!]
	\centering
	\includegraphics[width=0.9\linewidth]{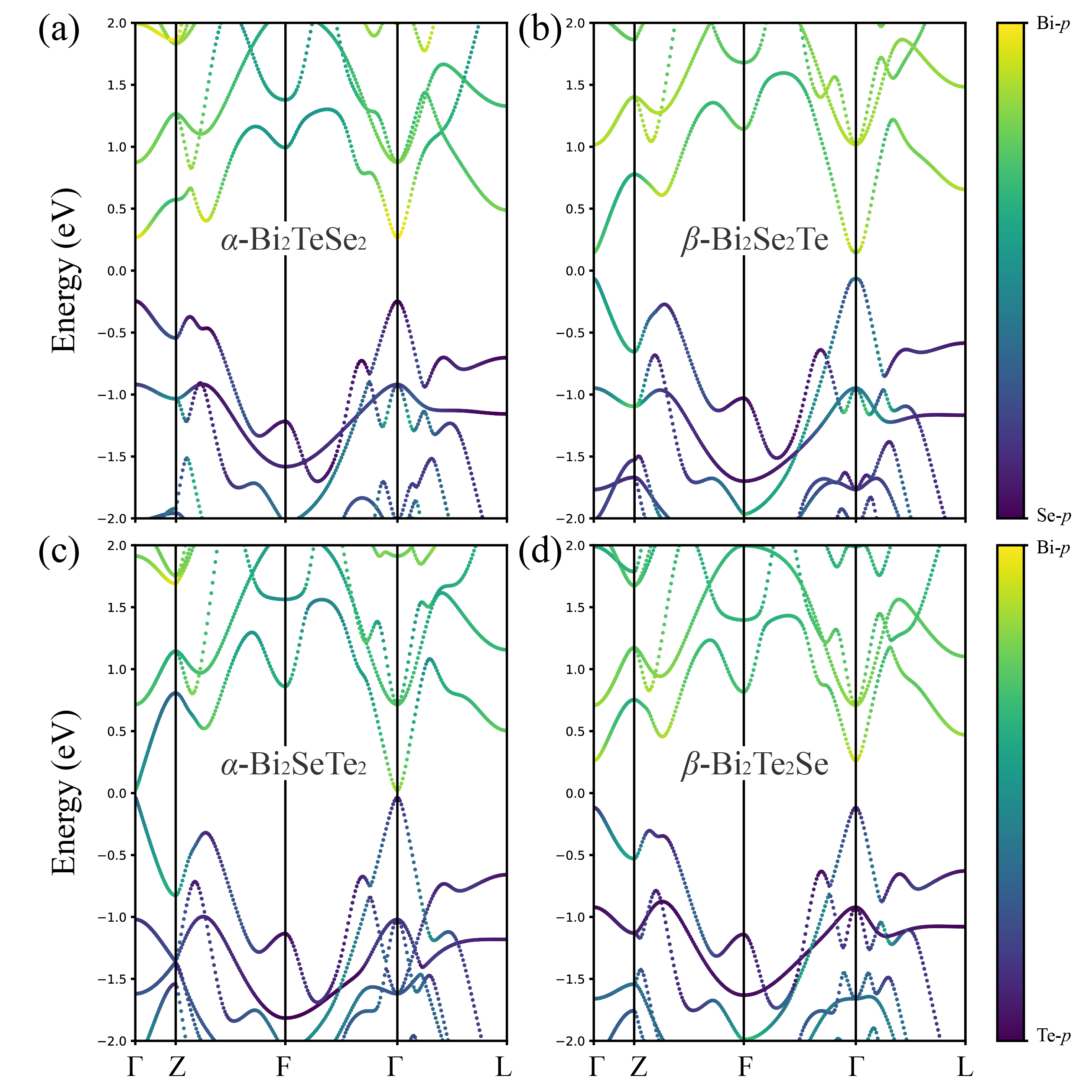}
	\caption{Electronic band structures of (a) $\alpha$-Bi$_2$TeSe$_2$, (b) $\beta$-Bi$_2$Se$_2$Te, (c) $\alpha$-Bi$_2$SeTe$_2$, and (d) $\beta$-Bi$_2$Te$_2$Se without SOC effect. The contributions of the $p$-orbitals from Bi and Se/Te atoms are projected onto the bands.}
	\label{fig:03-band}
\end{figure}

\begin{figure}[ht!]
	\centering
	\includegraphics[width=0.9\linewidth]{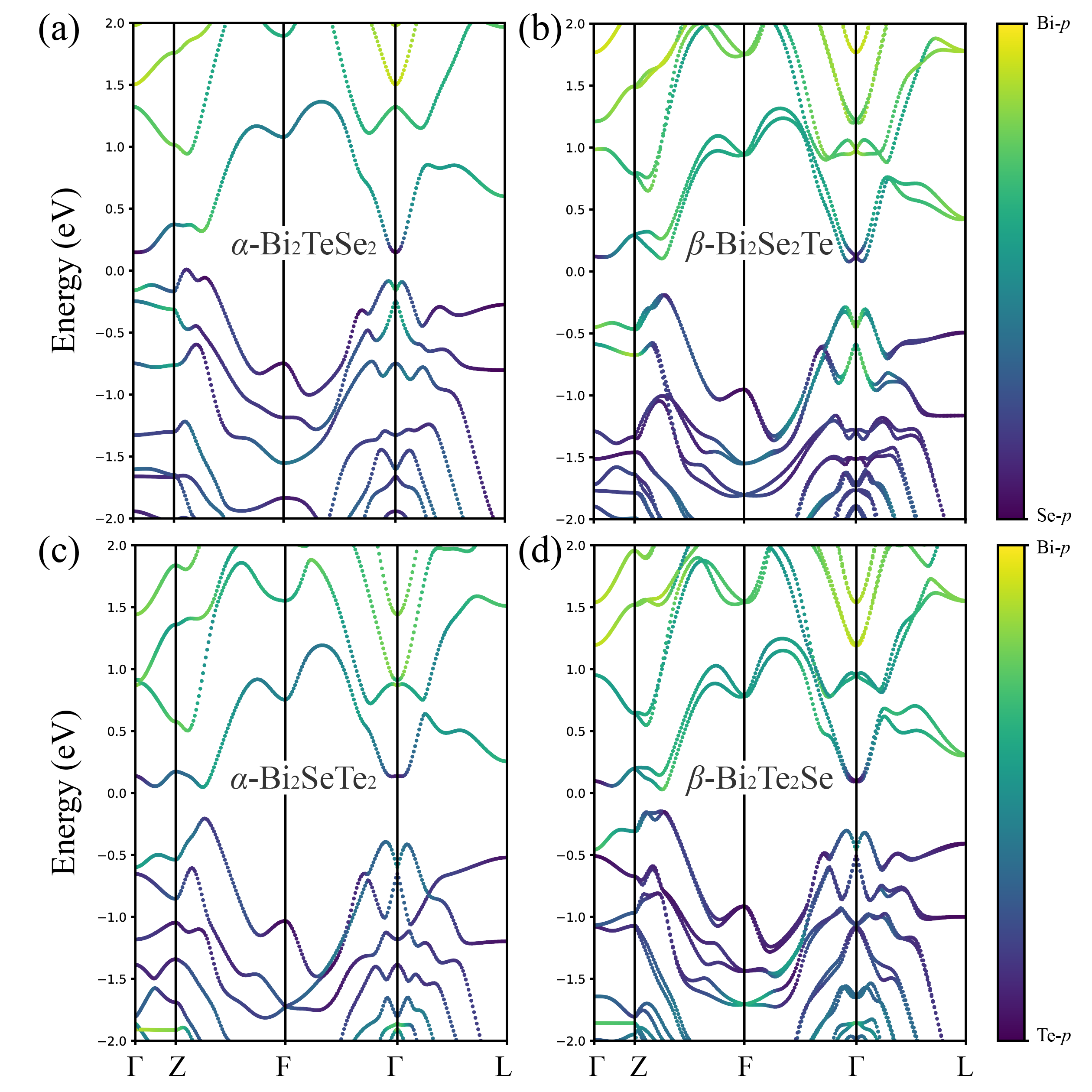}
	\caption{Electronic band structures of (a) $\alpha$-Bi$_2$TeSe$_2$, (b) $\beta$-Bi$_2$Se$_2$Te, (c) $\alpha$-Bi$_2$SeTe$_2$, and (d) $\beta$-Bi$_2$Te$_2$Se with SOC effect. The contributions of the $p$-orbitals from Bi and Se/Te atoms are projected onto the bands.}
	\label{fig:04-band}
\end{figure}

The orbital-projected band structures of Bi$_2$X$_2$Y along high-symmetry directions, calculated without and with SOC effect, are shown in Figs.~\ref{fig:03-band} and~\ref{fig:04-band}, respectively. Without SOC, all Bi$_2$X$_2$Y compounds exhibit a direct bandgap at the $\Gamma$ point, with the bandgap values $E^{DFT}_{g}$ being 0.51 eV for $\alpha$-Bi$_2$TeSe$_2$, 0.21 eV for $\beta$-Bi$_2$Se$_2$Te, 0.06 eV for $\alpha$-Bi$_2$SeTe$_2$, and 0.38 eV for $\beta$-Bi$_2$Te$_2$Se. The conduction band minimum (CBM) and valence band maximum (VBM) are primarily contributed by the $p$-orbitals of Bi and Se atoms, respectively. Apart from differences in the relative energy levels, the overall band structures of the four materials are remarkably similar. Although $\alpha$- and $\beta$-phases differ in symmetry, with the latter exhibiting broken inversion symmetry, the crystal field splitting of the orbitals remains consistent under both the D$_{3d}$ and C$_{3v}$ point groups. Consequently, the band dispersion trends are largely similar in the absence of SOC.

Consideration of SOC leads to marked changes in the electronic structure along high-symmetry paths, as shown in Fig.~\ref{fig:04-band}. The preservation of spin degeneracy in $\alpha$-phases is attributed to the combined protection of time-reversal symmetry $T$ ($\hat T: E_\uparrow({\bf{k}}) = E_\downarrow(-{\bf{k}})$) and spatial inversion symmetry $I$ ($\hat I: E_\uparrow({\bf{k}}) = E_\uparrow(-{\bf{k}})$), with no spin splitting observed in their band structures. In contrast, $\beta$-phases, which lack spatial inversion symmetry, exhibit varying degrees of band splitting along specific high-symmetry paths.  Notably, a pronounced Rashba splitting is observed at the $\Gamma$ point for $\beta$-Bi$_2$Se$_2$Te. The Rashba splitting can be described by the Hamiltonian $H_R = \alpha_R \sigma\left(k_{\|} \times E_Z\right)$, where $E_Z$ represents the out-of-plane electric field. This effect arises from the broken inversion symmetry in the Janus structure, which induces a spontaneous out-of-plane polarization field. Such a mechanism is advantageous for maintaining a low chemical potential at high carrier concentrations,\cite{wu_two-dimensional_2014} thereby probably enhancing the thermoelectric performance. 
Furthermore, all Bi$_2$X$_2$Y exhibit band inversion between the $p$-orbitals of Bi and Se/Te atoms at the $\Gamma$ point, accompanied by varying degrees of saddle-shaped band dispersion, which is a hallmark feature of topological insulators.\cite{liu_model_2010,zhang_topological_2009} Due to the heavier atomic mass of Te compared to Se, the relativistic effects are more pronounced in Te-containing compounds, leading to larger band inversion and wider bandgap at the $\Gamma$ point. These significant changes in the band structure result in bandgap values $E^{DFT+SOC}_{g}$ of 0.14 eV for $\alpha$-Bi$_2$TeSe$_2$, 0.27 eV for $\beta$-Bi$_2$Se$_2$Te, 0.25 eV for $\alpha$-Bi$_2$SeTe$_2$, and 0.17 eV for $\beta$-Bi$_2$Te$_2$Se along the high-symmetry paths.

To further verify the topological properties of Bi$_2$X$_2$Y, the topological invariants were calculated using the Wilson loop method. This approach is particularly suitable for $\beta$-phases with broken spatial inversion symmetry. The Wannier functions were constructed from the $p_{xyz}$ orbitals of Bi$_2$X$_2$Y. For all four Bi$_2$X$_2$Y materials, the evolution curves in the $k_i$ ($i=1, 2, 3$) = 0 plane cross the reference line an odd number of times (Fig. S2), corresponding to $Z_2$ invariant of 1. In contrast, in the $k_i$ ($i=1, 2, 3$) = 0.5 plane, the evolution curves cross the reference line an even number of times, yielding $Z_2$ invariant of 0. The topological invariants for all four materials are determined to be 1; (000), confirming that they are topological insulators.

\begin{figure}[ht!]
	\centering
	\includegraphics[width=0.85\linewidth]{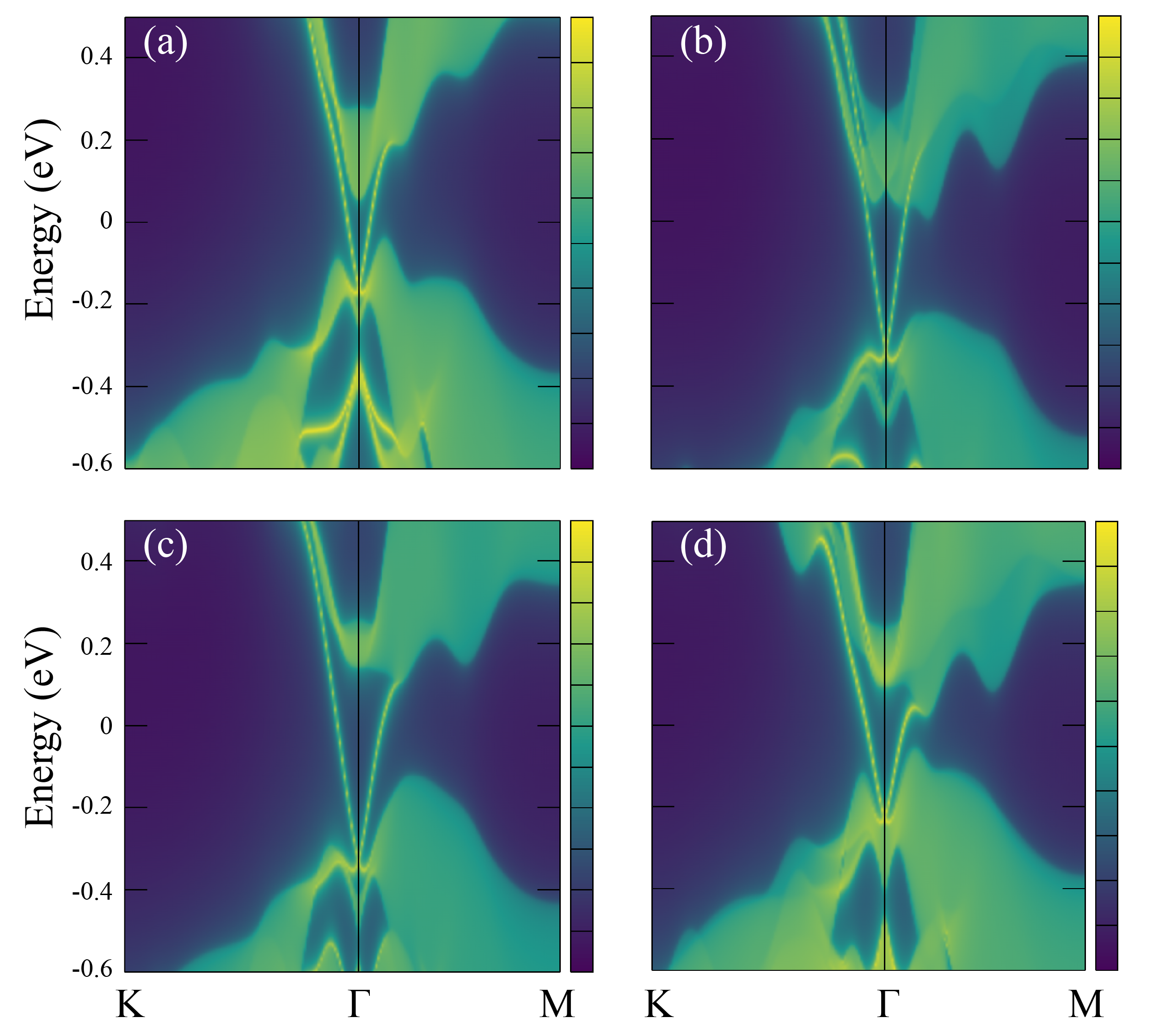}
	\caption{Topological boundary states on the (111) surface for (a) $\alpha$-Bi$_2$TeSe$_2$, (b) $\beta$-Bi$_2$Se$_2$Te, (c) $\alpha$-Bi$_2$SeTe$_2$, and (d) $\beta$-Bi$_2$Te$_2$Se.}
	\label{fig:05-surface}
\end{figure}

One of the hallmark features of topological insulators is their metallic surface states. Using the Green's function method implemented in WannierTools, the surface states of Bi$_2$X$_2$Y were calculated. As shown in Fig.~\ref{fig:05-surface}, all Bi$_2$X$_2$Y compounds exhibit distinct topological surface states, with conductive edge states emerging within the energy range of the bulk bandgap, bridging the conduction and valence bands. This provides further evidence for the topologically nontrivial nature of Bi$_2$X$_2$Y. It is worth noting that for $\beta$-phases, the two terminations on the (111) surface exhibit different surface states due to their distinct configurations, resulting in states at different energy depths. However, both terminations retain metallic properties (only one termination is shown in Fig.~\ref{fig:05-surface}). The unique characteristics of these surface states are expected to lead to diverse edge conduction properties, holding potential for future applications in electronic devices.

To gain deeper insight into the formation of the nontrivial electronic bands in Bi$_2$X$_2$Y, the low-energy effective Hamiltonian were derived to study the band structure near the Fermi level at the $\Gamma$ point. Although conventional topological insulators such as Bi$_2$Se$_3$, which share the same symmetry as $\alpha$-phases, are typically described using a 4-band model for their low-energy effective Hamiltonian,\cite{liu_model_2010} the sub valence band in Bi$_2$X$_2$Y exhibits strong coupling with the valence band due to their close energy proximity. Additionally, considering the $R3m$ symmetry of $\beta$-phases, a 6-band model (accounting for spin degeneracy at the $\Gamma$ point) was constructed to include the sub valence band, valence band, and conduction band for generality. For $\alpha$-Bi$_2$TeSe$_2$ and $\alpha$-Bi$_2$SeTe$_2$, the irreducible representations of the energy levels correspond to $\Gamma_5^-$ and $\Gamma_6^-$ (sub valence band), $\Gamma_4^+$ (valence band), and $\Gamma_4^-$ (conduction band) under the $D_{3d}$ double point group. The basis functions were transformed into the $\left| {J{M_J}} \right\rangle$ basis:
\begin{equation}
	\left|\frac{3}{2}^{-},-\frac{3}{2}\right\rangle-i\left|\frac{3}{2}^{-}, \frac{3}{2}\right\rangle \&-\left|\frac{3}{2}^{-}, \frac{3}{2}\right\rangle+i\left|\frac{3}{2}^{-},-\frac{3}{2}\right\rangle,\left|\frac{1}{2}, \pm \frac{1}{2}\right\rangle,\left|\frac{1}{2}, \pm \frac{1}{2}\right\rangle
\end{equation}
where the signs before and after the basis states denote parity and spin, respectively. For $\beta$-Bi$_2$Se$_2$Te and $\beta$-Bi$_2$Te$_2$Se, the irreducible representations of the sub valence band, valence band, and conduction band correspond to $\Gamma_6$, $\Gamma_4$, and $\Gamma_4$, respectively, under the $C_{3v}$ double point group. These representations are associated with the $\left| {J{M_J}} \right\rangle$ basis states:
\begin{equation}
-\left|\frac{3}{2}, \frac{3}{2}\right\rangle + i\left|\frac{3}{2},-\frac{3}{2}\right\rangle, \quad \left|\frac{1}{2}, \pm \frac{1}{2}\right\rangle, \quad \left|\frac{1}{2}, \pm \frac{1}{2}\right\rangle.
\end{equation}
Using these basis functions, the $\mathbf{k} \cdot \mathbf{p}$ model within a finite wave vector range can be constructed based on invariant theory.

For $\alpha$-phases, the relevant symmetry operations include spatial inversion symmetry, time-reversal symmetry, threefold rotational symmetry about the $z$-axis, and twofold rotational symmetry about the $x$-axis. The constrained symmetry operations for $\beta$-phases are time-reversal symmetry, threefold rotational symmetry about the $z$-axis, and mirror symmetry about the principal axis. By applying these symmetry operations to the basis states, the representation of the symmetry operations can be expressed as $\hat R\left| {J{M_J}} \right\rangle = D(\hat R)\left| {J{M_J}} \right\rangle$. Based on invariant theory, retaining terms up to second order in $k$ yields a 6-band low-energy effective $\mathbf{k} \cdot \mathbf{p}$ model near the $\Gamma$ point. By fitting the parameters to the DFT-calculated band structures, the coefficients of the low-energy effective Hamiltonian matrix were determined. Table~\ref{tab:table2} lists the coefficients ${M_i}$, ${B_{zi}}$, and ${B_{\bot i}}$ for the unperturbed diagonal terms of the Hamiltonian matrix for Bi$_2$X$_2$Y, expressed as $M({\bf{k}}) = {M_i} + {B_{zi}}k_z^2 + {B_{\bot i}}k_\bot^2$, where $i=1, 2, 3$ correspond to the conduction band, valence band, and subvalence band, respectively, and $k_\bot^2 = k_x^2 + k_y^2$. For Bi$_2$X$_2$Y, the eigenenergies of the conduction and valence bands at the $\Gamma$ point (${\bf{k}}=0$) are given by ${M_1}$ and ${M_2}$, respectively. The coefficients ${B_{z1}}$ (${B_{\bot 1}}$) for the conduction band are negative, while ${B_{z2}}$ (${B_{\bot 2}}$) for the valence band are positive, leading to band inversion at finite ${\bf{k}}$, similar to the low-energy effective model of Bi$_2$Se$_3$.\cite{zhang_topological_2009} This Hamiltonian accurately describes the topological properties of Bi$_2$X$_2$Y. Figure S3 compares the band structures obtained from the $\mathbf{k} \cdot \mathbf{p}$ model and DFT calculations, showing excellent agreement near ${\bf{k}}=0$.

\begin{table}[h!]
    \centering
    \caption{Coefficients of the unperturbed diagonal terms in the low-energy effective Hamiltonian for Bi$_2$X$_2$Y. The parameters $M_i$ (in eV), $B_{zi}$ (in eV nm$^2$), and $B_{\bot i}$ (in eV nm$^2$) correspond to the conduction band ($i=1$), valence band ($i=2$), and subvalence band ($i=3$), respectively.}
    \setlength{\tabcolsep}{3mm}{
    \begin{tabular}{cccccccccc}
    \hline
    Materials & $M_1$ & $B_{z1}$ & $B_{\bot 1}$ & $M_2$ & $B_{z2}$ & $B_{\bot 2}$ & $M_3$ & $B_{z3}$ & $B_{\bot 3}$ \\ \hline
    $\alpha$-Bi$_2$TeSe$_2$ & 0.149 & -0.012 & -0.105 & -0.158 & 0.087 & 0.148 & -0.245 & -0.003 & -0.069 \\
    $\beta$-Bi$_2$Se$_2$Te & 0.122 & -0.036 & -0.149 & -0.448 & 0.085 & 0.135 & -0.586 & -0.006 & -0.047 \\
    $\alpha$-Bi$_2$SeTe$_2$ & 0.136 & -0.086 & -0.189 & -0.599 & 0.108 & 0.077 & -0.654 & -0.023 & -0.092 \\
    $\beta$-Bi$_2$Te$_2$Se & 0.096 & -0.036 & -0.148 & -0.458 & 0.111 & 0.096 & -0.510 & -0.020 & -0.091 \\ \hline
    \end{tabular}}
	\label{tab:table2}
\end{table}

Through electronic structure calculations and model construction, it has been confirmed that Bi$_2$X$_2$Y compounds are topological insulators. The interplay between band inversion, a hallmark of their topological nature, and Rashba splitting induced by symmetry breaking in the Janus structure results in a rich multivalley feature near the Fermi level in Bi$_2$X$_2$Y. This suggests an enhanced phase space for intravalley and intervalley carrier scattering in Bi$_2$X$_2$Y, which could have significant implications for their electronic and transport properties.

All Bi$_2$X$_2$Y compounds are identified as topological insulators with varying degrees of band inversion. As previously discussed, the band inversion induces saddle-shaped band dispersion, which enhances valley degeneracy near the Fermi level, thereby facilitating band convergence. This effect is particularly significant in $\beta$-Bi$_2$Se$_2$Te, where the saddle-shaped dispersion lies in close proximity to both the valence band maximum and the conduction band minimum, leading to improved electrical transport properties at low doping concentrations. Moreover, the $\beta$-Bi$_2$Se$_2$Te exhibits pronounced Rashba splitting, which further enhances band convergence by splitting the valleys, offering the potential for further improvements in electrical transport performance. Additionally, the intrinsic Rashba effect in these heavy-element-based compounds may generate an internal polarization field, increasing bond anharmonicity and reducing lattice thermal conductivity.\cite{wu_two-dimensional_2014} Consequently, the novel Janus $\beta$-phases are expected to exhibit superior thermoelectric performance compared to their prototype counterparts.

\subsection{Electrical transport properties}

\begin{figure}[ht!]
	\centering
	\includegraphics[width=0.9\linewidth]{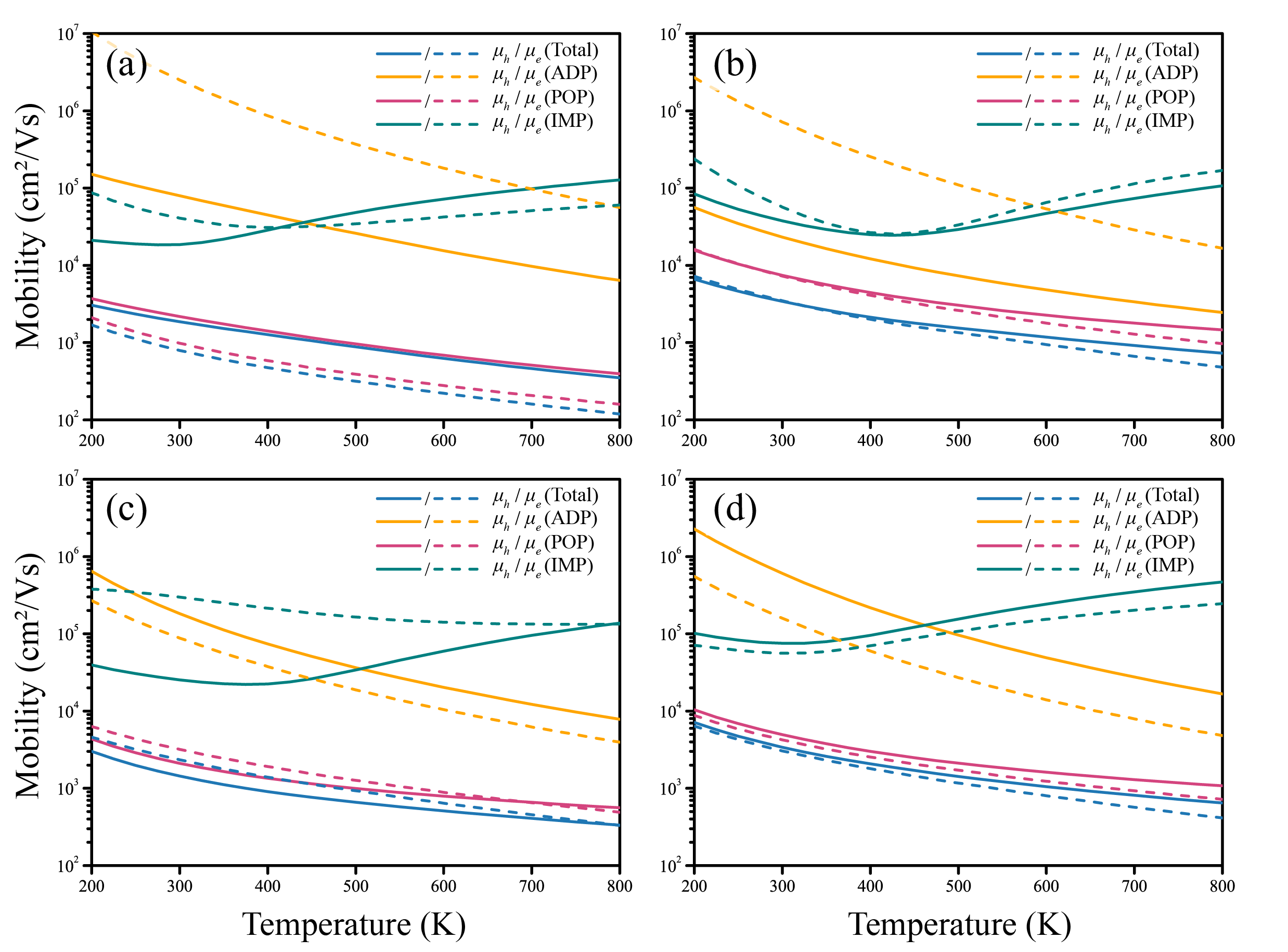}
	\caption{Carrier mobilities of (a) $\alpha$-Bi$_2$TeSe$_2$, (b) $\beta$-Bi$_2$Se$_2$Te, (c) $\alpha$-Bi$_2$SeTe$_2$, and (d) $\beta$-Bi$_2$Te$_2$Se at the doping concentration of $1 \times 10^{18}$ cm$^{-3}$ as a function of temperature.}
	\label{fig:06-mobility}
\end{figure}

On the basis of the electronic structure including SOC effect, the carrier mobility of Bi$_2$X$_2$Y at the doping concentration of $1 \times 10^{18}$ cm$^{-3}$ was calculated using the Boltzmann transport equation within the momentum relaxation time approximation (Fig.~\ref{fig:06-mobility}). The calculations accounted for both elastic and inelastic scattering mechanisms, with elastic scattering including ADP scattering and IMP scattering, and inelastic scattering dominated by POP scattering. At room temperature (300 K), the electron (hole) mobilities for $\alpha$-Bi$_2$TeSe$_2$, $\beta$-Bi$_2$Se$_2$Te, $\alpha$-Bi$_2$SeTe$_2$, and $\beta$-Bi$_2$Te$_2$Se were found to be 794 (1860), 3482 (3419), 2338 (1439), and 3056 (3403) cm$^2$/Vs, respectively, demonstrating excellent electrical properties. Notably, $\beta$-phases exhibit significantly higher mobilities compared to their counterparts.  
While the carrier mobilities exhibit variations among the materials, polar optical phonon scattering emerges as the predominant mechanism governing both electron and hole transport characteristics in all four compounds. This finding reveals that the conventional deformation potential formalism fails to adequately characterize their scattering mechanisms.
Over a wide temperature range, $\alpha$-Bi$_2$TeSe$_2$ shows a notably lower electron mobility compared to its hole mobility, while the other three materials exhibit comparable electron and hole mobilities, with $\alpha$-Bi$_2$SeTe$_2$ displaying slightly higher electron mobility. This behavior can be preliminarily explained by the band structure, as $\alpha$-Bi$_2$TeSe$_2$ exhibits a simpler dispersion at the conduction band edge compared to the other materials. Furthermore, it is noteworthy that acoustic deformation potential scattering contributes significantly to the mobility in $\beta$-Bi$_2$Se$_2$Te, and ionized impurity scattering also plays a notable role at low temperatures, a feature not observed in the other three materials.

\begin{figure}[ht!]
	\centering
	\includegraphics[width=0.9\linewidth]{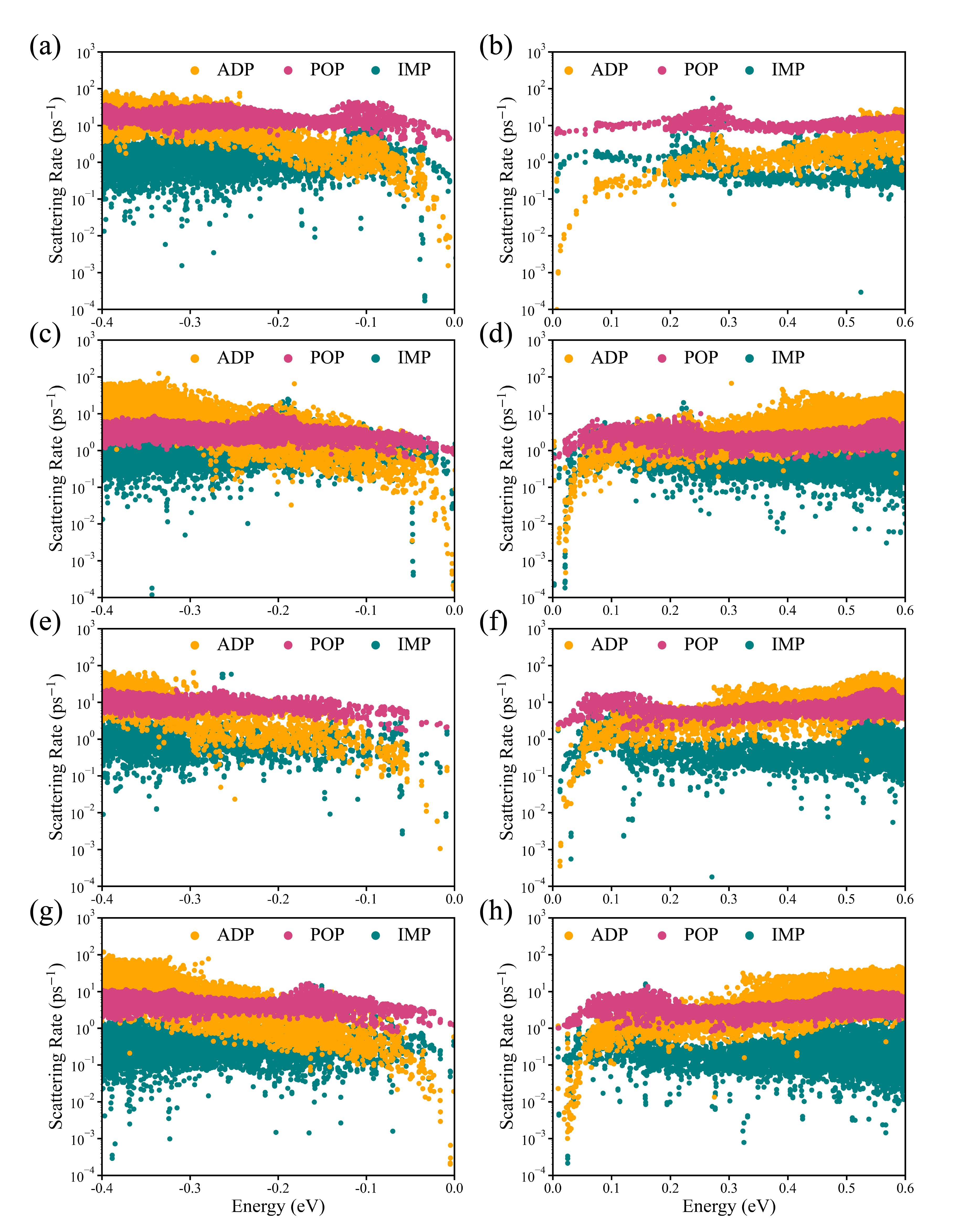}
	\caption{Hole and electron scattering rates for (a, b) $\alpha$-Bi$_2$TeSe$_2$, (c, d) $\beta$-Bi$_2$Se$_2$Te, (e, f) $\alpha$-Bi$_2$SeTe$_2$, and (g, h) $\beta$-Bi$_2$Te$_2$Se at room temperature and the doping concentration of $1 \times 10^{18}$ cm$^{-3}$, respectively.}
	\label{fig:07-scattering}
\end{figure}

To further elucidate the variation in carrier mobility and uncover the underlying scattering mechanisms in Bi$_2$X$_2$Y, the scattering-rate-resolved contributions for holes and electrons at the band edges were calculated at 300 K and the doping concentration of $1 \times 10^{18}$ cm$^{-3}$ (Fig.~\ref{fig:07-scattering}). It is evident that POP scattering dominates the scattering rates at the band edges, playing the most significant role in determining carrier mobility. By comparing the scattering rates at the band edges across the four materials, the relative magnitudes of their mobilities can be inferred. $\beta$-phases exhibit lower scattering rates at the band edges compared to $\alpha$-phases, consistent with their higher mobilities. Furthermore, while the ADP and IMP scattering rates are generally similar in magnitude across the materials, the POP scattering rates differ significantly. The enhanced mobilities of $\beta$-phases can be primarily attributed to their reduced POP scattering rates, which are influenced by both the phonon dispersion and the material's polarity. Additionally, the variations in scattering rates are closely related to differences in the electronic band dispersion. Notably, the electron scattering rates of $\beta$-Bi$_2$Se$_2$Te, $\alpha$-Bi$_2$SeTe$_2$, and $\beta$-Bi$_2$Te$_2$Se exhibit a pronounced increase within a small energy range (~0.05 eV), corresponding to the presence of degenerate valleys in the band structure, which enhances the scattering probability. In contrast, $\alpha$-Bi$_2$TeSe$_2$, which lacks significant multivalley features in its conduction band, shows an increase in electron scattering rates around 0.2 eV, associated with the emergence of small valleys along the Z-F high-symmetry direction. A similar analysis applies to the hole scattering rates.

\begin{figure}[ht!]
	\centering
	\includegraphics[width=0.9\linewidth]{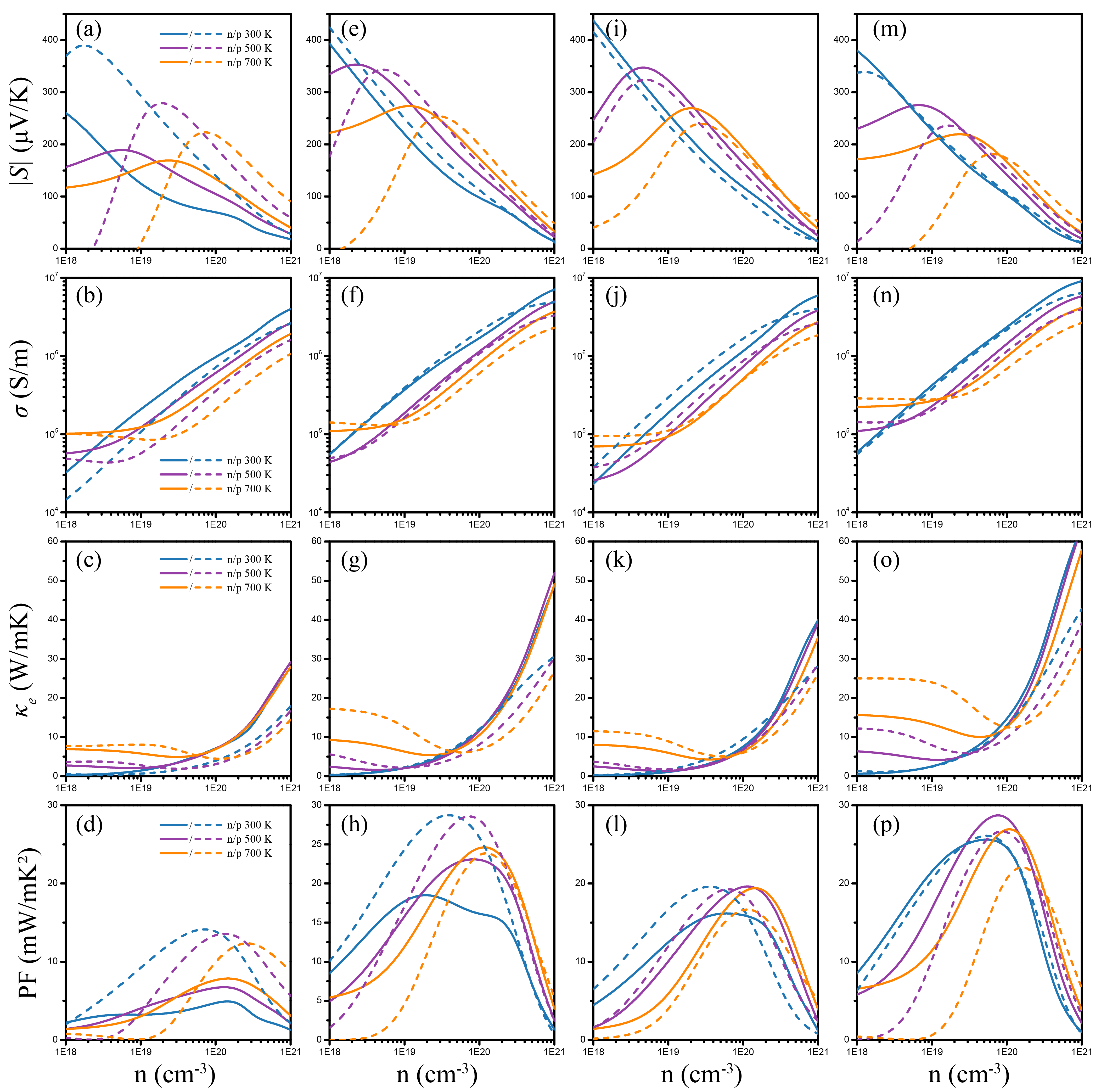}
	\caption{The Seebeck coefficient (a, e, i, m), electrical conductivity (b, f, j, n), electronic thermal conductivity (c, g, k, o), and power factor (d, h, l, p) as functions of carrier concentration for $\alpha$-Bi$_2$TeSe$_2$, $\beta$-Bi$_2$Se$_2$Te, $\alpha$-Bi$_2$SeTe$_2$, and $\beta$-Bi$_2$Te$_2$Se, respectively.}
	\label{fig:08-transport}
\end{figure}

Based on the aforementioned scattering mechanisms, the carrier relaxation times were determined, and the Boltzmann transport equation was solved to obtain the absolute Seebeck coefficient $|S|$, electrical conductivity $\sigma$, electronic thermal conductivity $\kappa_{e}$, and power factor PF of Bi$_2$X$_2$Y as functions of carrier concentration and temperature (Fig.~\ref{fig:08-transport}). The Seebeck coefficient can be described by the Mott relation,\cite{sun_large_2015} expressed as,
\begin{equation}
	S = -\frac{\pi^2}{3} \frac{k_{\mathrm{B}}^2 T}{e} \left[ \frac{\partial \ln \tau(E)}{\partial E} + \frac{\partial \ln N(E)}{\partial E} \right]_{E_{\mathrm{f}}},
\end{equation}
where $\tau(E)$ and $N(E)$ represent the energy-dependent carrier relaxation time and density of states (DOS), respectively. The terms inside the brackets are divided into a scattering-related term and a band-related term. In the conventional constant relaxation time approximation (CRTA) method, the electron relaxation time is assumed to be constant, resulting in a scattering term of zero. In this work, however, the energy dependence of the carrier relaxation time is explicitly considered, providing a more accurate description of the thermoelectric properties.

The Seebeck coefficients of all four materials exhibit a characteristic trend under p-type doping: an initial increase followed by a decrease with increasing carrier concentration. This behavior is indicative of bipolar conduction, a phenomenon commonly observed in narrow-band gap semiconductors at elevated temperatures.\cite{parker_thermoelectric_2012,wang_band_2021} At low levels of p-type doping, hole carriers predominantly govern transport properties. However, at elevated temperatures, the contribution of thermally excited electrons to the Seebeck coefficient becomes non-negligible. As the hole concentration further increases, the influence of electrons gradually diminishes. Bipolar conduction is detrimental to thermoelectric performance, and among the materials, $\beta$-Bi$_2$Se$_2$Te, with its relatively larger bandgap compared to its counterparts, exhibits the weakest bipolar conduction. In contrast, $\beta$-Bi$_2$Te$_2$Se, which has a smaller bandgap compared to $\alpha$-Bi$_2$SeTe$_2$, shows pronounced bipolar conduction even at high doping concentrations. Additionally, due to the Rashba splitting effect, $\beta$-Bi$_2$Se$_2$Te achieves band convergence, resulting in higher Seebeck coefficients than $\alpha$-Bi$_2$TeSe$_2$ under both p-type and n-type doping. In contrast, $\beta$-Bi$_2$Te$_2$Se, which lacks significant Rashba splitting and has a reduced bandgap compared to $\alpha$-Bi$_2$SeTe$_2$, exhibits strong bipolar conduction and overall lower Seebeck coefficients. Regarding electrical conductivity and electronic thermal conductivity, both are proportional to carrier mobility, and $\beta$-phases demonstrate higher values than their counterparts. The calculated power factors reveal a significant enhancement for $\beta$-Bi$_2$Se$_2$Te compared to $\alpha$-Bi$_2$TeSe$_2$. Although the Seebeck coefficient of $\beta$-Bi$_2$Te$_2$Se is suppressed, its superior electrical conductivity results in a higher power factor than that of $\alpha$-Bi$_2$SeTe$_2$.

\subsection{Thermal transport properties}

\begin{figure}[ht!]
	\centering
	\includegraphics[width=0.9\linewidth]{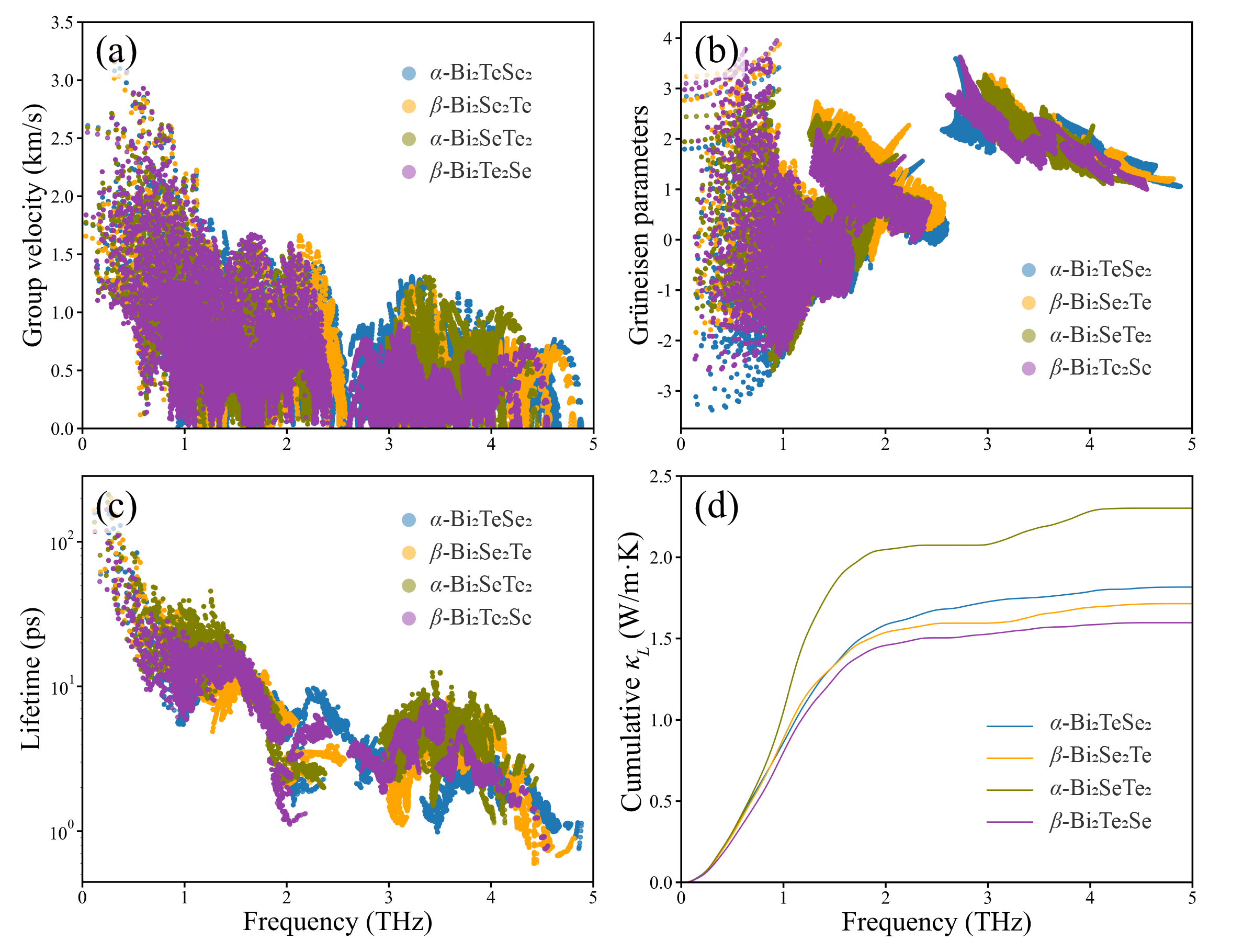}
	\caption{(a) Phonon group velocity, (b) Grüneisen parameter, (c) phonon lifetime, and (d) cumulative lattice thermal conductivity of $\alpha$-Bi$_2$TeSe$_2$, $\beta$-Bi$_2$Se$_2$Te, $\alpha$-Bi$_2$SeTe$_2$, and $\beta$-Bi$_2$Te$_2$Se at 300 K.}
	\label{fig:09-thermal}
\end{figure}

The force constants of Bi$_2$X$_2$Y at finite temperatures were obtained using the temperature-dependent effective potential (TDEP) method. These were subsequently incorporated into the phonon Boltzmann transport framework to investigate phonon scattering and related thermal transport properties. By sampling the force constants at 300 K and renormalizing the phonon dispersion, the phonon group velocities of Bi$_2$X$_2$Y were determined, as shown in Fig.~\ref{fig:09-thermal}a. The group velocities of the four materials exhibit similar magnitudes overall. Notably, the group velocities of optical phonons near 2 THz are significantly higher than those of high-frequency optical phonons. Moreover, their frequencies are close to those of acoustic phonons, facilitating efficient acoustic-optical phonon scattering. In particular, $\beta$-Bi$_2$Te$_2$Se exhibits noticeably lower group velocities for high-frequency optical phonons compared to the other three materials, which can be attributed to the relatively flat and localized phonon modes in the high-frequency region of its renormalized phonon dispersion, a feature also observed in its ground-state phonon dispersion.  

To characterize the anharmonic interactions in Bi$_2$X$_2$Y, the Grüneisen parameter was analyzed (Fig.~\ref{fig:09-thermal}b). Materials with weak anharmonicity generally have Grüneisen parameters below 1, while strongly anharmonic systems, such as PbTe and BiCuSeO, typically exceed this threshold.\cite{morelli_intrinsically_2008,pei_high_2013} 
The results indicate that a significant number of phonon modes in Bi$_2$X$_2$Y exhibit Grüneisen parameters exceeding 1 across the entire frequency range, particularly for both acoustic and high-frequency optical phonons. 
This pronounced anharmonicity, evidenced by the large Grüneisen parameters, leads to vigorous phonon-phonon scattering, which in turn shortens phonon relaxation times and suppresses the thermal transport contribution from high-frequency optical phonons.

The phonon lifetimes of Bi$_2$X$_2$Y at 300 K were determined through third-order force constant calculations at finite temperatures (Fig.~\ref{fig:09-thermal}c). The lifetimes of acoustic and low-frequency optical phonons fall within the range of 10-100 ps, whereas high-frequency optical phonons exhibit significantly shorter lifetimes, ranging from 1 to 10 ps, corresponding to larger phonon linewidths. For compositions with identical elemental constituents, such as $\alpha$-Bi$_2$SeTe$_2$ and $\beta$-Bi$_2$Te$_2$Se, the latter exhibits a higher phonon scattering rate (inverse lifetime) in the low-frequency region. 
This can be attributed to two key features in the phonon dispersion of $\alpha$--Bi$_2$SeTe$_2$: the presence of a distinct phonon band gap between low- and high-frequency optical modes, and a notable lack of avoided crossing between the acoustic and optical branches. The absence of this strong acoustic-optical mode coupling significantly reduces the available phase space for scattering, leading to weaker phonon interactions and consequently longer lifetimes compared to the other three materials where such crossings are prominent.

Based on the phonon scattering properties, the lattice thermal conductivity ($\kappa_{L}$) of Bi$_2$X$_2$Y was computed. The cumulative $\kappa_{L}$ as a function of frequency are shown in Fig. 9(d), where it is evident that phonons below 2 THz, including acoustic and low-frequency optical modes, contribute predominantly to thermal transport, consistent with the aforementioned analysis. The weak thermal accumulation at high frequencies in $\beta$-Bi$_2$Se$_2$Te and $\alpha$-Bi$_2$SeTe$_2$ is attributed to localized high-frequency phonon states. Considering finite-temperature phonon scattering, the in-plane $\kappa_{L}$ of $\alpha$-Bi$_2$TeSe$_2$, $\beta$-Bi$_2$Se$_2$Te, $\alpha$-Bi$_2$SeTe$_2$, and $\beta$-Bi$_2$Te$_2$Se at room temperature are 2.28, 2.16, 2.72, and 1.93 W·m$^{-1}$·K$^{-1}$, respectively, while the corresponding out-of-plane values are 0.90, 0.81, 1.46, and 0.92 W·m$^{-1}$·K$^{-1}$. The lower out-of-plane thermal conductivity stems from the weak van der Waals interactions between the quintuple layers in Bi$_2$X$_2$Y.

\begin{figure}[ht!]
	\centering
	\includegraphics[width=0.7\linewidth]{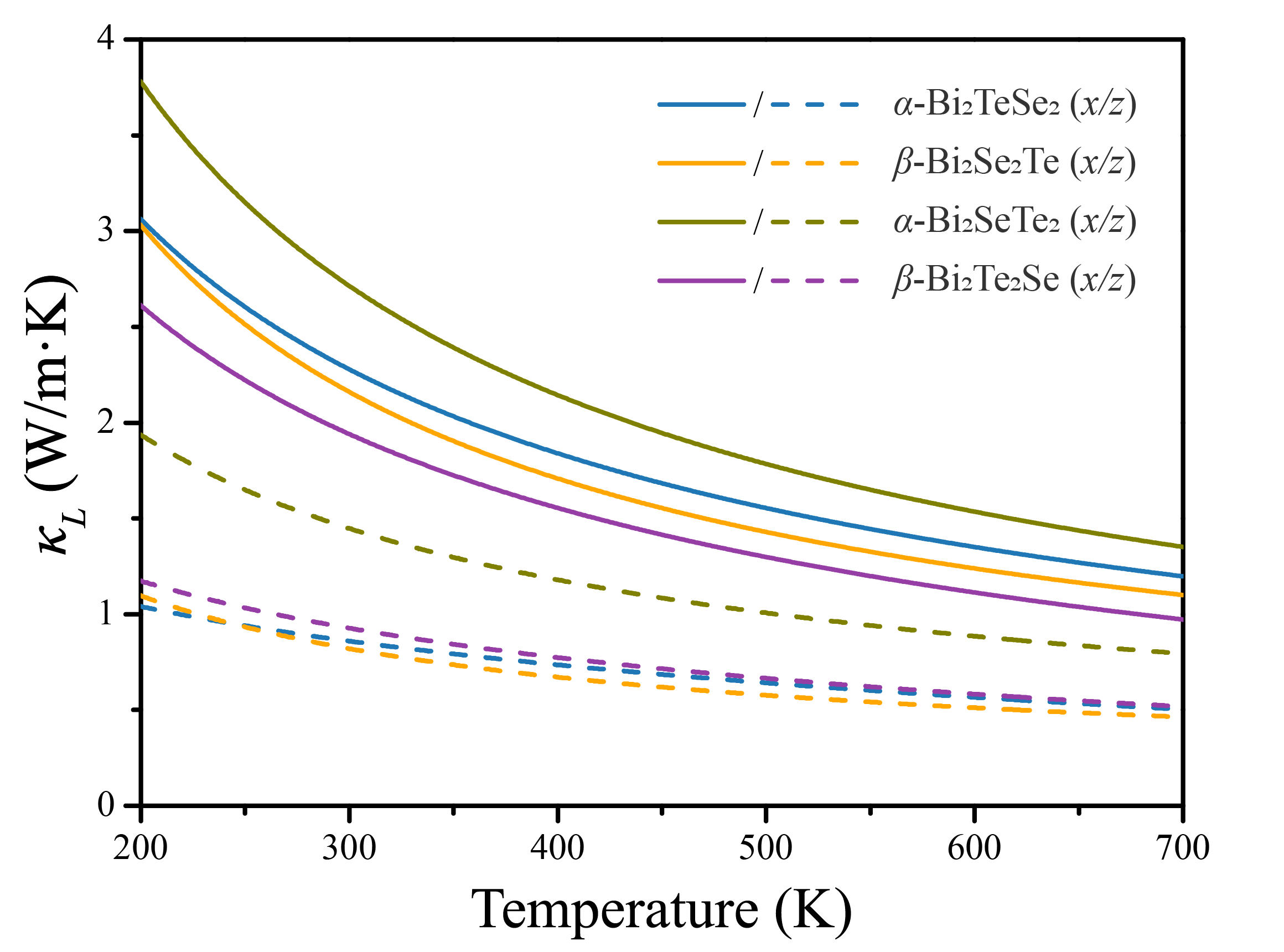}
	\caption{Temperature-dependent lattice thermal conductivity of (a) $\alpha$-Bi$_2$TeSe$_2$, (b) $\beta$-Bi$_2$Se$_2$Te, (c) $\alpha$-Bi$_2$SeTe$_2$, and (d) $\beta$-Bi$_2$Te$_2$Se in the in-plane and out-of-plane directions.}
	\label{fig:10-thermal}
\end{figure}

The temperature-dependent $\kappa_{L}$ of Bi$_2$X$_2$Y was calculated, and the results were fitted to analyze their behavior, as shown in Fig.~\ref{fig:10-thermal}. The results indicate that in the mid-to-high temperature range, Bi$_2$X$_2$Y exhibits similar $\kappa_{L}$ values and temperature dependence. The temperature dependence of the out-of-plane $\kappa_{L}$ is slightly weaker than that of the in-plane $\kappa_{L}$, a trend that has also been reported in previous studies.\cite{cen_structural_2021} Across the entire temperature range, $\alpha$-Bi$_2$SeTe$_2$ exhibits the highest lattice thermal conductivity in both directions, a finding consistent with its weaker overall phonon scattering as discussed above. Notably, the ranking of in-plane and out-of-plane thermal conductivities for $\beta$--Bi$_2$Te$_2$Se is inconsistent with the other materials, which may be attributed to unique anisotropic scattering channels arising from the strong structural polarity along its out-of-plane direction. Overall, all four Bi$_2$X$_2$Y compounds exhibit low lattice thermal conductivity over a broad temperature range. Crucially, the inversion-symmetry-breaking $\beta$-phases demonstrate lower average thermal conductivities compared to their centrosymmetric $\alpha$-phase counterparts. The Janus structure effectively suppresses lattice thermal conductivity, which is beneficial for enhancing thermoelectric performance.

\subsection{Thermoelectric performance}

\begin{figure}[ht!]
	\centering
	\includegraphics[width=0.9\linewidth]{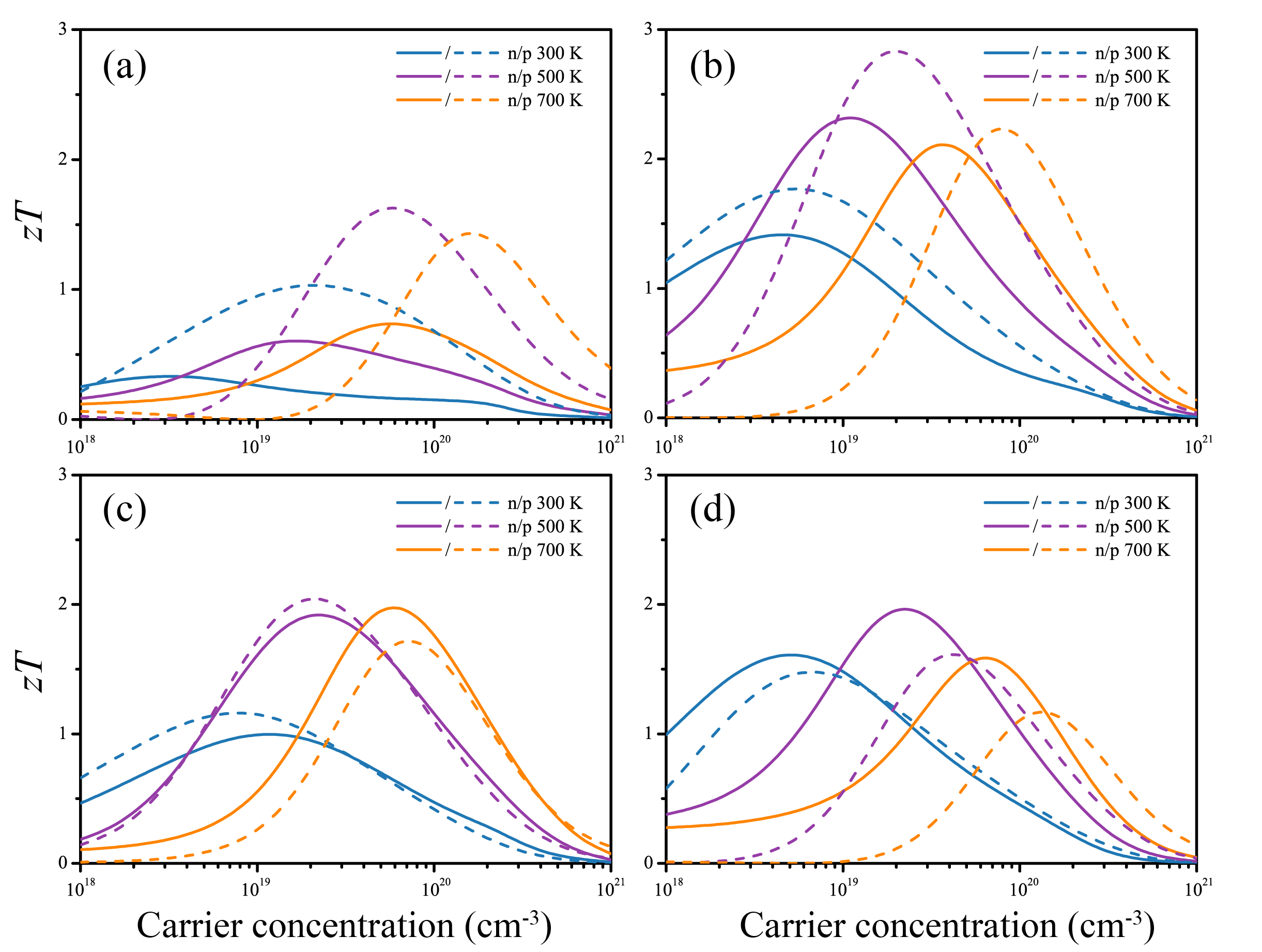}
	\caption{Thermoelectric figure of merit as a function of carrier concentration at different temperatures for (a) $\alpha$-Bi$_2$TeSe$_2$, (b) $\beta$-Bi$_2$Se$_2$Te, (c) $\alpha$-Bi$_2$SeTe$_2$, and (d) $\beta$-Bi$_2$Te$_2$Se.}
	\label{fig:11-thermoelectric}
\end{figure}

Based on the electrical and thermal transport properties of Bi$_2$X$_2$Y compounds, the thermoelectric performance can be evaluated by calculating the figure of merit ($zT = S^2\sigma T / (\kappa_e + \kappa_L)$). Fig.~\ref{fig:11-thermoelectric} presents the variation of $zT$ values for four materials as a function of $n$-type and $p$-type doping concentrations across different temperatures. Owing to its superior electrical transport properties and low lattice thermal conductivity, $\beta$-Bi$_2$Se$_2$Te exhibits $zT$ values exceeding 1 over a wide concentration range in the medium-to-high temperature regime, with a peak value of 2.82 achieved at 500 K under $p$-type doping concentration of $2 \times 10^{19}~\text{cm}^{-3}$. This represents a significant enhancement in thermoelectric performance from $\alpha$-Bi$_2$TeSe$_2$ to $\beta$-Bi$_2$Se$_2$Te. Although $\beta$-Bi$_2$Te$_2$Se possesses a high power factor and nearly minimal lattice thermal conductivity across a broad temperature range, its thermoelectric performance improvement over $\alpha$-Bi$_2$SeTe$_2$ is not as pronounced due to its elevated electronic thermal conductivity. Nevertheless, $\beta$-Bi$_2$Te$_2$Se remains a promising thermoelectric material candidate. It should be noted that while the electrical and thermal transport parameters of Bi$_2$X$_2$Y were obtained within reasonable approximations, the quantitative values primarily serve as references to experimental data, such as the $zT$ value around 1 for Bi$_2$(Te, Se)$_3$ alloys.\cite{pei_bi2te3-based_2020-1} More importantly, theoretical insights into enhancing thermoelectric performance were achieved in this work. Specifically, the designed novel Janus structure breaks spatial inversion symmetry, introducing Rashba splitting while maintaining the topological properties of band inversion, thereby realizing band convergence enhancement. 

\section{Conclusion}

In conclusion, the thermoelectric properties of computationally designed Janus materials, $\beta$-Bi$_2$Se$_2$Te and $\beta$-Bi$_2$Te$_2$Se, have been systematically investigated using first-principles calculations. It is revealed that the broken spatial inversion symmetry in these Janus structures induces a significant Rashba splitting. This effect, acting in synergy with the intrinsic topological band inversion, promotes a high degree of band convergence near the Fermi level. The engineered band degeneracy simultaneously enhances the power factor via an increased Seebeck coefficient and suppresses the lattice thermal conductivity due to increased phonon scattering. This synergistic optimization of electronic and phononic transport culminates in a high thermoelectric figure of merit, with a peak $zT$ value predicted to exceed 1 for $\beta$-Bi$_2$Se$_2$Te in the medium-to-high temperature range. This work not only identifies $\beta$-phase bismuth-based compounds as promising high-performance thermoelectric candidates but, more importantly, establishes a novel and effective design principle: harnessing the intrinsic spin-orbit coupling effects within topological materials to rationally engineer band structures for superior thermoelectric performance.

\begin{acknowledgments}
	This work is supported by the National Natural Science Foundation of China (62275053, 61775042, 11674062, 11374063, 12304038), Shanghai Municipal Natural Science Foundation under Grant Nos. 19ZR1402900 and the Fudan University-CIOMP Joint Fund (FC2017-003).
\end{acknowledgments}

\section*{REFERENCES}

\bibliography{Reference}

\end{document}


\title{Supplementary Information\\
Synergy of Rashba and Topological Effects for High-Performance Bismuth-Based Thermoelectrics}

\author{Lei Peng}
\affiliation{School of Information Science and Technology, Department of Optical Science and Engineering, Fudan University, Shanghai 200433, China}
\affiliation{Key Laboratory for Computational Physical Sciences (MOE), Institute of Computational Physical Sciences and Department of Physics, Fudan University, Shanghai 200433, China}

\author{Ruixiao Lian}
\affiliation{School of Information Science and Technology, Department of Optical Science and Engineering, Fudan University, Shanghai 200433, China}

\author{Hongyu Chen}
\affiliation{School of Information Science and Technology, Department of Optical Science and Engineering, Fudan University, Shanghai 200433, China}

\author{Ben Li}
\affiliation{School of Information Science and Technology, Department of Optical Science and Engineering, Fudan University, Shanghai 200433, China}

\author{Yu Wu}
\affiliation{Advanced Thermal Management Technology and Functional Materials Laboratory, Ministry of Education Key Laboratory of NSLSCS, School of Energy and Mechanical Engineering, Nanjing Normal University, Nanjing 210023, P. R. China}

\author{Yuxiang Zheng} \email{yxzheng@fudan.edu.cn}
\affiliation{School of Information Science and Technology, Department of Optical Science and Engineering, Fudan University, Shanghai 200433, China}
\affiliation{Yiwu Research Institute of Fudan University, Yiwu City, Zhejiang 322000, China}

\author{Hao Zhang} \email{zhangh@fudan.edu.cn}
\affiliation{School of Information Science and Technology, Department of Optical Science and Engineering, Fudan University, Shanghai 200433, China}
\affiliation{Yiwu Research Institute of Fudan University, Yiwu City, Zhejiang 322000, China}

\maketitle

\newpage
\begin{table}[h!]
    \centering
    \caption{Elastic constants (C$_{ij}$, GPa), bulk modulus (B, GPa), shear modulus (G, GPa), and Young’s modulus (Y, GPa) of Bi$_2$X$_2$Y.}
    \setlength{\tabcolsep}{2.5mm}{
    \begin{tabular}{cccccccccc}
    \hline
    Materials               & C$_{11}$ & C$_{33}$ & C$_{44}$ & C$_{12}$ & C$_{13}$ & C$_{14}$  & B     & G     & Y     \\ \hline
    $\alpha$-Bi$_2$TeSe$_2$ & 80.937 & 45.352 & 28.719 & 25.066 & 25.644 & -14.481 & 39.99 & 25.80 & 63.70 \\
    $\beta$-Bi$_2$Se$_2$Te  & 83.271 & 45.602 & 26.859 & 23.423 & 25.917 & -13.654 & 40.30 & 25.85 & 63.90 \\
    $\alpha$-Bi$_2$SeTe$_2$ & 82.145 & 50.715 & 30.472 & 22.402 & 28.849 & -15.863 & 41.69 & 27.16 & 66.94 \\
    $\beta$-Bi$_2$Te$_2$Se  & 79.833 & 49.973 & 32.088 & 23.149 & 28.025 & -16.446 & 40.89 & 27.20 & 66.79 \\ \hline
    \end{tabular}}
\end{table}

\newpage

\begin{figure}[ht!]
\centering
\includegraphics[width=0.9\linewidth]{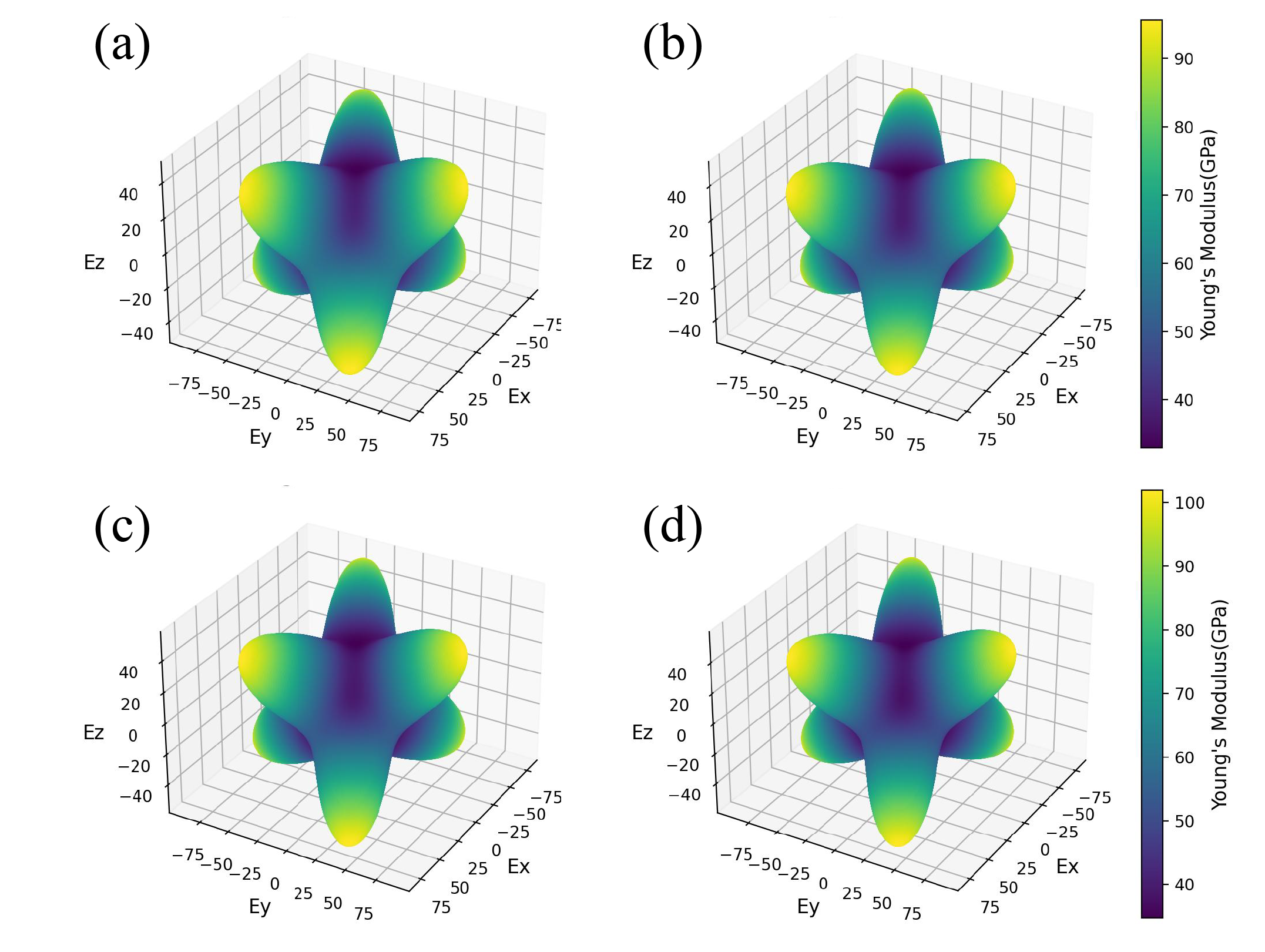}
\caption{The three-dimensional spatial distributions of Young’s modulus for (a) $\alpha$-Bi$_2$TeSe$_2$, (b) $\beta$-Bi$_2$Se$_2$Te, (c) $\alpha$-Bi$_2$SeTe$_2$, and (d) $\beta$-Bi$_2$Te$_2$Se.}
\label{figs:Y3D} 
\end{figure}

\begin{figure}[ht!]
    \centering
    \includegraphics[width=0.9\linewidth]{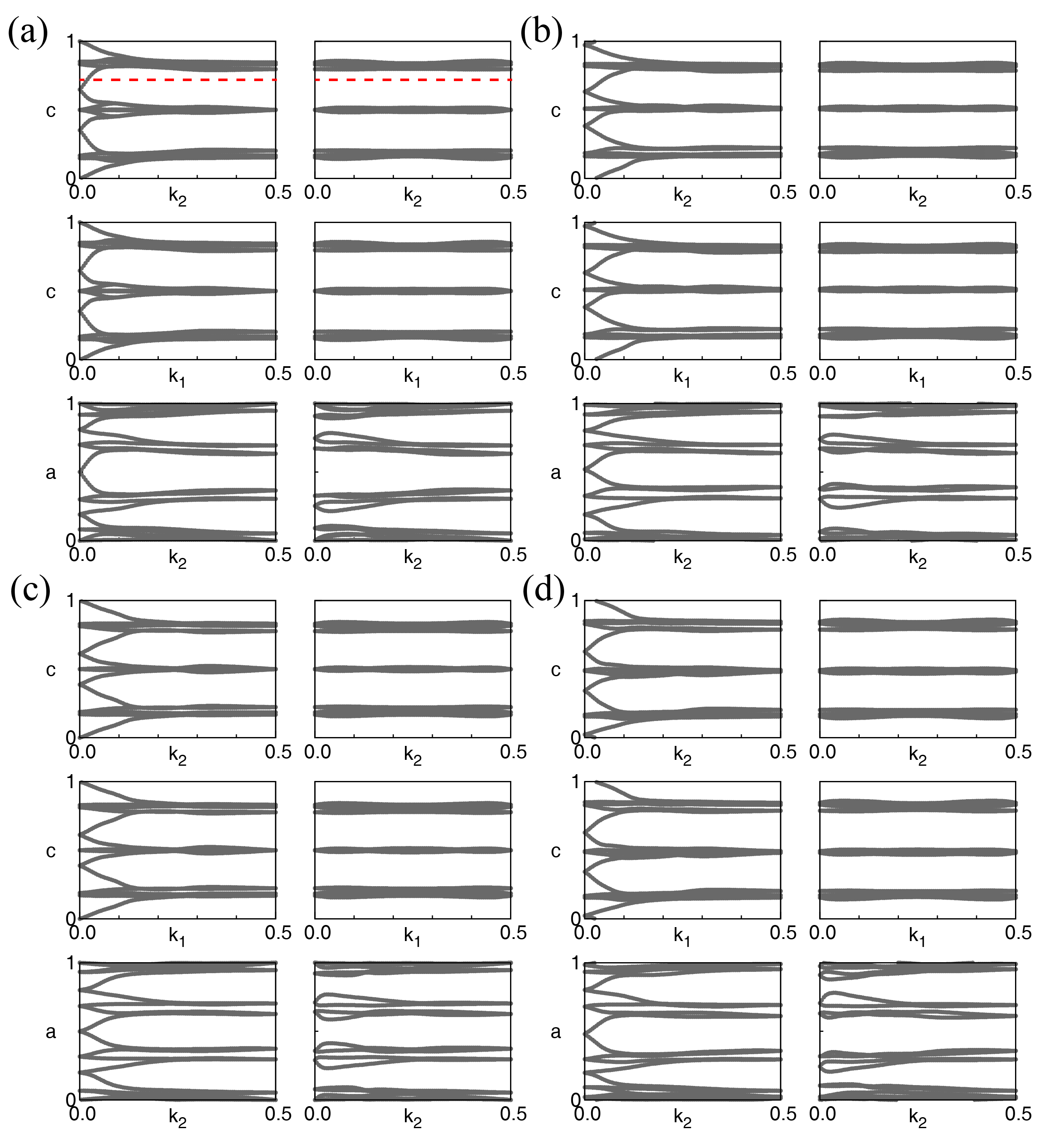}
    \caption{Evolution of the Wannier function centers for (a) $\alpha$-Bi$_2$TeSe$_2$, (b) $\beta$-Bi$_2$Se$_2$Te, (c) $\alpha$-Bi$_2$SeTe$_2$, and (d) $\beta$-Bi$_2$Te$_2$Se. The red dashed line represents the reference line.}
    \label{figs:WCC} 
\end{figure}

\begin{figure}[ht!]
    \centering
    \includegraphics[width=0.8\linewidth]{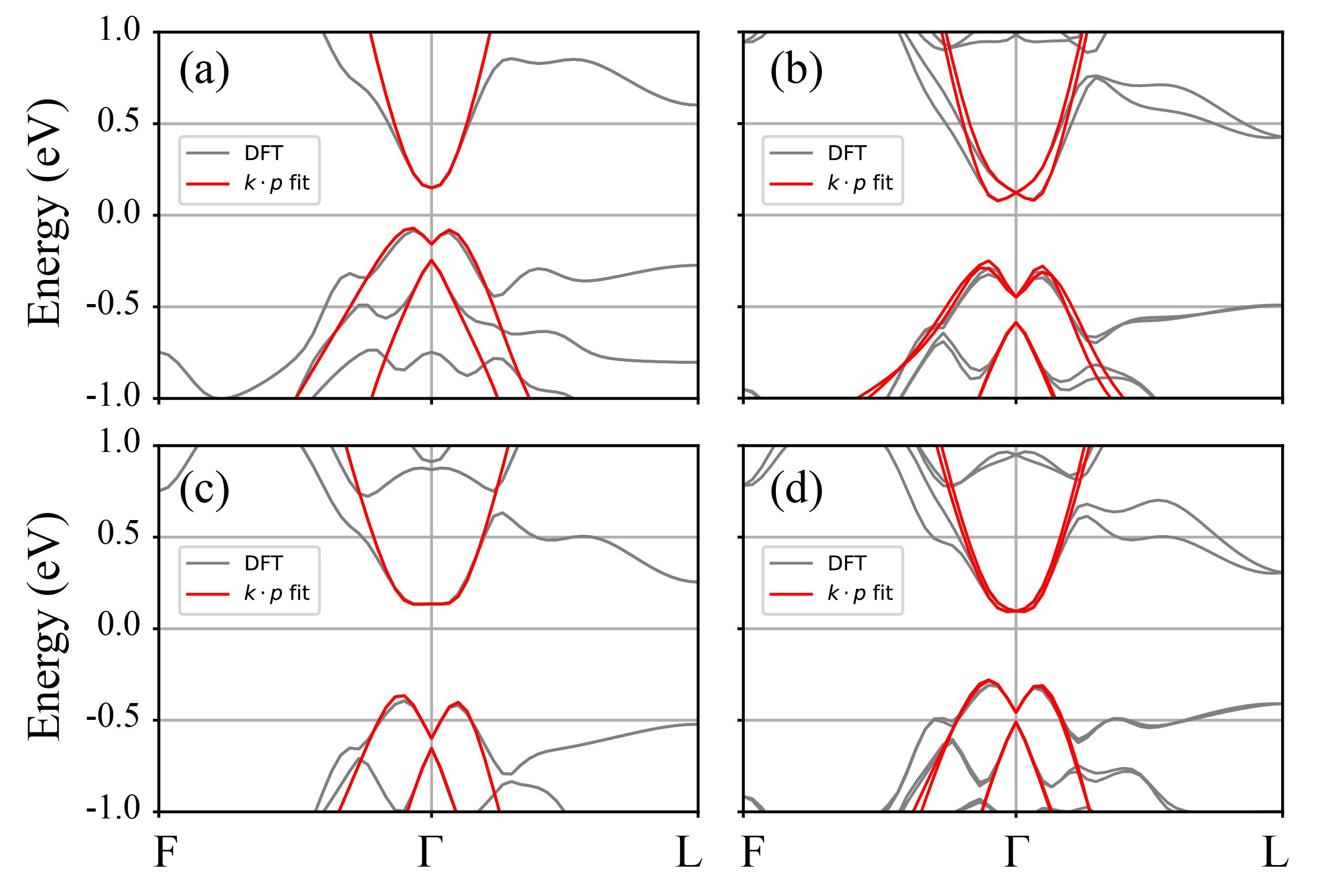}
    \caption{Comparison of the band structures obtained from the $k \cdot p$ model (red curves) and DFT calculations (gray curves) for (a) $\alpha$-Bi$_2$TeSe$_2$, (b) $\beta$-Bi$_2$Se$_2$Te, (c) $\alpha$-Bi$_2$SeTe$_2$, and (d) $\beta$-Bi$_2$Te$_2$Se.}
    \label{figs:kpband} 
\end{figure}